%% file: main.tex
\documentclass[10pt, conference, letterpaper]{IEEEtran} 

\AtBeginDocument{%
  \providecommand\BibTeX{{%
    \normalfont B\kern-0.5em{\scshape i\kern-0.25em b}\kern-0.8em\TeX}}}
    
\usepackage[T1]{fontenc}
\usepackage{cite}
\usepackage{tabularx}
\usepackage{colortbl}
\usepackage{graphicx}
\usepackage{float}
\graphicspath{ {figs/} }
\usepackage{dsfont}
\usepackage{amsmath}
 
\usepackage[ruled,vlined]{algorithm2e}
\usepackage[font=small]{caption}
\usepackage[font=scriptsize]{subcaption}
\usepackage[export]{adjustbox}
\usepackage{siunitx}
\usepackage{colortbl}
\usepackage{algorithmic}
\usepackage[thinlines]{easytable}
\usepackage{siunitx}
\usepackage[normalem]{ulem}
\usepackage{amsfonts}
\usepackage{multirow,verbatim}
\usepackage[normalem]{ulem}
\usepackage[dvipsnames]{xcolor}
\usepackage{url}
\usepackage{soul}
\usepackage{tikz}
\usetikzlibrary{tikzmark}
\usepackage{multirow}
\usepackage{amsthm}
\usepackage[english]{babel}
\usepackage{xcolor}
\colorlet{blue}{black}
\usepackage{algorithmic}

\usepackage{booktabs,makecell,multirow,hhline,threeparttable} 
\usepackage{xcolor,colortbl,pgf}

\newif\ifcleanversion

\cleanversionfalse  

\ifcleanversion
    \colorlet{blue}{black}
\fi

\definecolor{Color1}{RGB}{240, 240, 240}
\usepackage{makecell}
\usepackage[top=0.7in, bottom=1.0in, left=0.7in, right=0.7in]{geometry}

\definecolor{lightgray}{RGB}{211,211,211}
\definecolor{LightCyan}{RGB}{224,255,255}

\newcommand{\name}{\text{PRISM}}

\SetKwComment{Comment}{$\triangleright$\ }{}

\newcolumntype{b}{>{\columncolor{blue!10}}c}
\newcolumntype{y}{>{\columncolor{yellow!10}}c}
\newcolumntype{d}{>{\columncolor{red!7}}c}

\SetKwInput{KwInput}{Input}                
\SetKwInput{KwOutput}{Output}              

\begin{document}

\title{PRISM: Propagation-based RF Indoor Sensing and Material Detection using Mobile Transmitter{
\footnotesize \textsuperscript{}}
\thanks{*Wentao Gao and Sunday Amatare have equally contributed to this work.}}

\author{	\IEEEauthorblockN{
\small
       Wentao Gao\IEEEauthorrefmark{1},  Sunday Amatare\IEEEauthorrefmark{1},  Xiaojun Shang, and
        Debashri Roy
        }
\IEEEauthorblockA{
\small
Department of Computer Science and Engineering, The University of Texas at Arlington\\
Emails: {\tt \{wxg8464, saa3326\}@mavs.uta.edu, \{xiaojun.shang, debashri.roy\}@uta.edu} \\
\IEEEauthorrefmark{1}{\em Wentao Gao and Sunday Amatare have equally contributed to this work}
}
}

\maketitle


\begin{abstract}

Accurately locating objects and identifying their materials indoors is key to emerging mobile computing applications, including 6G networked robotics, mobile healthcare, pervasive security, and mobile edge computing (MEC)-enabled services. Camera-based methods raise privacy concerns and fail in non-line-of-sight (NLoS) conditions, limiting their use at the edge. Radio frequency (RF) propagation maps, paired with a digital twin (DT), offer a promising alternative, but prior work assumes a fixed transmitter location, relies on heavy vision-style models, and treats material recognition as a closed-set problem, ill-suited for the real world.
We present \name{} (\textbf{P}ropagation-based \textbf{R}F \textbf{I}ndoor \textbf{S}ensing and \textbf{M}aterial detection), a lightweight, edge-friendly framework that addresses these gaps. {\color{blue}\name~leverages a Vision Transformer (ViT) backbone paired with a Detection Transformer (DETR)-style set-prediction head. Rather than simply treating RF maps as proxy images, we exploit the structural alignment between Transformer attention mechanisms and RF physics: the global self-attention natively captures the long-range, continuous dependencies of multipath fading better than localized convolutional filters. Furthermore, \name~introduces three advances: (i) robustness to mobile transmitters via location-invariant feature learning; (ii) fast adaptation to new environments through few-shot fine-tuning with as few as $25$ samples, and (iii) native open-set anomaly rejection by re-engineering the bipartite matching loss to force out-of-distribution materials into the background class.}
Evaluated on Blender-based digital twins with NVIDIA Sionna RT ray tracing, \name{} achieves an overall F1 of 0.8802 under the default inference configuration and 0.9822 after applying a class-specific confidence threshold selected exclusively on an independent validation subset  across concrete, glass, metal, and wood, and rejects $100\%$ of unknown materials. {\color{blue} A sim-to-real study further demonstrates that co-training
with eight real and 40 synthetic maps achieves correct
localization and material identification for all eight
objects in two held-out real maps, while maintaining
99.3\% detection accuracy on unseen synthetic scenarios.} These results establish \name{} as a practical, privacy-preserving, and efficient solution for RF-based indoor sensing in resource-constrained, MEC-driven deployments.
\end{abstract}


\input{Sections/introduction-new}
\input{Sections/related-work-new}

\input{Sections/problem-formulation-new}
\input{Sections/proposed-framework-new}
\input{Sections/dataset-new}
\input{Sections/experiments-new}
\input{Sections/evaluation-new}
\input{Sections/8-conclusion}

\bibliographystyle{IEEEtran} 
\bibliography{reference}

\end{document}

%% file: Sections/introduction-new.tex
\section{Introduction}
\label{sec:intro}
\vspace*{-3pt}

Accurate localization and characterization of objects within an environment are essential for a wide range of mission-critical applications, spanning autonomous robotics, industrial manufacturing, smart healthcare, and advanced security systems. These applications are increasingly supported by mobile edge computing (MEC) infrastructures, where sensing, inference, and decision-making are performed close to end users to meet stringent latency and reliability requirements \cite{mao2017survey}. Traditionally, machine learning (ML) has leveraged large-scale, camera-based image datasets to achieve precise object localization and semantic characterization \cite{deng2009imagenet,redmon2016you,zohar2023prob}.

\begin{figure}[t!]
\centering

\includegraphics[width=0.85\linewidth]{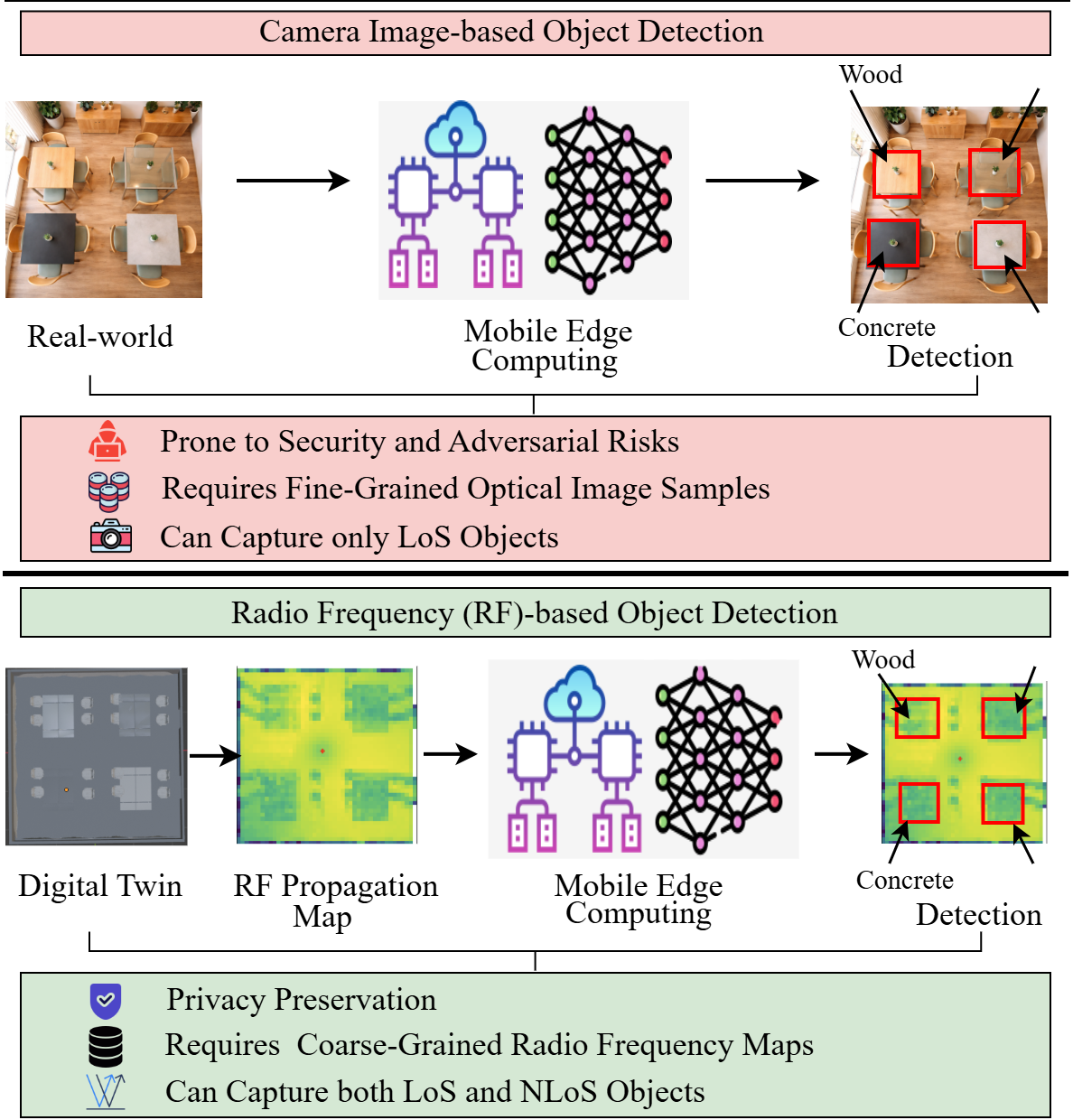}

\caption{Comparison of camera image-based and RF-based object detection. We demonstrate that \name~is privacy-preserving, robust to non-line-of-sight conditions, and highly computationally efficient for edge deployment.}
\vspace{-20pt}
\label{fig:intro-fig}
\end{figure}

However, vision-based methods present significant operational and ethical challenges. They intrinsically divulge sensitive private information, making them unsuitable for deployment in residential or security-sensitive spaces. Furthermore, optical sensors are severely limited by environmental constraints; they fail to capture objects or in poor lighting, and/or non-line-of-sight (NLoS) conditions, i.e., settings in which no direct optical path exists between the camera and the target due to intervening obstacles such as walls or furniture. They also require extensive, manually labeled datasets that are highly vulnerable to adversarial risks and data poisoning. Their high computational and data demands further restrict efficient deployment in resource-constrained MEC environments.

To overcome these visual limitations, the wireless communications and sensing communities have increasingly turned to radio frequency (RF) signals, which enable robust object detection and perception under adverse and NLoS conditions \cite{lai2024enabling, ercan2024ris, woodford2023metasight}. Driven by the emergence of integrated sensing and communications (ISAC) in next-generation networks, RF sensing offers a unified framework for both communication and environmental perception. As a key enabler for 6G networked robotics and the mobile Internet of Things (IoT), ISAC provides continuous sensing intelligence at the network edge. Unlike optical waves, RF signals penetrate occlusions and are highly sensitive to the physical and electromagnetic properties of the environment. Because different materials possess distinct relative permittivity and conductivity, they interact with RF waves uniquely, causing distinct patterns of reflection, diffraction, and scattering. This physical phenomenon makes RF signals inherently well-suited for not merely locating an object, but explicitly characterizing its material composition.

Despite the theoretical advantages of RF sensing, acquiring large-scale, high-fidelity empirical RF data to train robust ML models is notoriously resource-intensive. To bridge this gap, recent literature has explored the integration of digital twins (DTs), deterministic ray-tracing, and neural radio-frequency radiance fields (NeRF2) to simulate and model RF propagation \cite{Nerf2, nerf2-channel, radiotwin}. These state-of-the-art approaches demonstrate the feasibility of combining RF simulation with ML within a DT environment to circumvent privacy concerns while characterizing objects' shapes and material compositions. Beyond static scene simulation, these environments can be continuously synchronized with real-time changes in object positions, providing a scalable source of training data that enables continuous adaptation for mobile IoT and MEC-enabled edge intelligence.

Despite these innovative premises, current state-of-the-art RF sensing frameworks exhibit notable architectural and operational limitations that hinder their practical, large-scale deployment. First, existing models demonstrate a profound sensitivity to the spatial configuration of the environment. They typically require centralized, static transmitter placements and struggle to generalize when real-world constraints force asymmetric, edge-bound, or highly mobile transmitter locations. Second, current methodologies heavily rely on fine-tuning pre-existing, vision-centric ML models (e.g., YOLO architectures). Applying off-the-shelf visual models to RF data overlooks a fundamental modality gap. Optical images exhibit dense spatial resolution with sharp object 
boundaries, whereas RF propagation maps capture the continuous, 
diffuse distribution of electromagnetic signal strength shaped by 
reflection, diffraction, and multipath fading. These two modalities 
therefore carry fundamentally different spatial statistics. 
Convolutional architectures, which are optimized to extract local 
edge and texture patterns and are known to be less effective at 
modeling long-range spatial dependencies~\cite{guo2022attention, 
dosovitskiy2021vit}, are not a natural fit for RF heatmaps, where 
discriminative information is distributed across broader spatial 
regions rather than concentrated along sharp edges. This observation 
is consistent with recent work on radio map estimation, which has 
shown that most state-of-the-art methods still rely on CNN backbones 
even though transformer-based architectures offer a better match to 
the global structure of RF data~\cite{wang2025rmtransformer, 
viet2024storm}.

Furthermore, {\color{black} beyond the architectural modality gap,} state-of-the-art systems primarily operate under a strict closed-set assumption. Real-world MEC deployments represent open-world environments where an object localization system will inevitably encounter materials absent from its offline training corpus. A traditional closed-set model forcefully and erroneously misclassifies these anomalies into known categories, critically undermining system reliability. Therefore, addressing open-set material detection with a lightweight, edge-native architecture is a strict prerequisite for autonomous operation in next-generation sensing systems.

{\color{black}
To overcome these combined limitations, we propose \name, a robust, edge-friendly localization framework (illustrated in Fig.~\ref{fig:intro-fig}). The core architectural novelty of \name~lies in recognizing and actively exploiting the structural alignment between Transformer attention mechanisms and the physics of RF wave propagation. While early cross-disciplinary works treated RF heatmaps as proxy images for standard CNNs, \name~shifts the paradigm. By replacing rigid convolutional anchors with global self-attention and dynamic set-prediction, the architecture natively captures the continuous, distributed nature of multipath interference. 

By leveraging the deterministic nature of DT ray-tracing to guarantee an infinitely scalable data source for robust training, this architectural shift enables \name~to solve the critical real-world challenges of transmitter mobility, domain shift, and open-world uncertainty.} 
Specifically, the primary contributions of this work are summarized as follows:

\begin{enumerate}

\item {\color{black}\textbf{Robustness to Mobile Transmitters via Location-Invariant Feature Learning:} We propose a novel feature extraction methodology that overcomes the transmitter location dependency prevalent in state-of-the-art models. By decoupling detection accuracy from the signal's spatial origin, $\name$ accurately localizes and characterizes materials regardless of the transmitter's coordinates, enabling robust operation for mobile networked agents.}


\item \textbf{Cross-Scenario Generalization via Few-Shot Adaptation:} We introduce a robust training-and-testing pipeline capable of extreme domain adaptation. $\name$, trained in a baseline environmental scenario, achieves high accuracy on entirely unseen, highly complex, and multi-path-rich benchmark data by utilizing only a minimal support set for rapid fine-tuning.

\item \textbf{Open-Set Material Detection and Anomaly Rejection:} We address the unpredictability of real-world indoor environments by equipping the model with an unknown rejection capability. {\color{black}By repurposing the bipartite matching algorithm to force anomalous scattering signatures into a \texttt{background} target, \name~effectively suppresses bounding-box outputs for out-of-distribution materials without relying on post-hoc gating networks, significantly reducing false positives in dynamic settings.}


\end{enumerate}

Moreover, the supporting contributions of this paper are listed as follows:
\begin{enumerate}

\item \textbf{Lightweight, Edge-Native Architecture:} We design an open-source ML architecture optimized specifically for the physical characteristics of RF propagation maps. By avoiding the retrofitting of heavy, vision-centric models, \name~drastically reduces computational overhead and inference latency, {\color{black} reducing computational overhead relative to vision-centric detectors and enabling rapid MEC-server-side adaptation.}

\item \textbf{Scalable Data Synthesis and Open-Source Release
{\color{black}with Real-World Validation}:}
By leveraging deterministic DT-based RF propagation synthesis,
we enable scalable data generation for domain adaptation.
{\color{black}We further collect 10 real-world propagation maps
and demonstrate that real-synthetic co-training improves
real-world detection while preserving generalization
to unseen synthetic scenarios.}
To foster further research, we release an open-source
dataset comprising $3{,}133$ high-quality synthesized
RF propagation maps {\color{black}and 10 measured maps},
alongside our digital twin models and complete
$\name$ codebase.\footnote{The dataset, digital twin
models, and model codebase are publicly available at
\url{https://github.com/Dextergao14/PRISM}.}

\item \textbf{Validation under Challenging RF Conditions:} We rigorously evaluate $\name$ against heavily curated ``hard samples,'' demonstrating its superior robustness, detection accuracy, and reliability in non-obvious, complex, and multi-path-dominant RF propagation environments.
\end{enumerate}

The remainder of this article is organized as follows. Section~\ref{sec:related_works} reviews the related work in machine learning and DT-enabled sensing. Section~\ref{sec:problem-formulation} details the system model and problem formulation, including the expanded open-set definitions. Section~\ref{sec:proposed_framework} presents the proposed \name~architecture and the lightweight ML model design. Section~\ref{sec:testbed_design} outlines the data synthesis pipeline and dataset. Section~\ref{sec:experiments} and \ref{sec:evaluation} provide a detailed discussion of experiments and a comprehensive performance evaluation. Finally, Section~\ref{sec:conclusions} concludes the paper.

%% file: Sections/related-work-new.tex
\section{Related Work}
\label{sec:related_works}

\noindent\textcolor{blue}{{\textbf{RF and WiFi Sensing.}} RF-based sensing has been widely explored for activity recognition~\cite{ma2019wifisurvey}, localization~\cite{kotaru2015spotfi}, occupancy estimation~\cite{depatla2015occupancy}, and environmental perception~\cite{zhao2018rfpose} using measurements such as RSSI, CSI, Doppler signatures, and channel impulse responses. These systems demonstrate that multipath propagation contains rich information about the surrounding physical environment~\cite{ma2019wifisurvey}. However, most RF sensing studies infer human activities, positions, or coarse environmental states from temporal channel observations rather than directly localizing physical structures and estimating their material composition from spatial propagation maps. PRISM considers the complementary problem of extracting object-level material information from the spatial distribution of RF propagation.}

\noindent\textcolor{blue}{\noindent{\bf Radar-Based Object and Material Sensing.} FMCW and mmWave radar systems provide fine-grained range, angle, and Doppler measurements and have been used for object detection and imaging~\cite{prabhakara2023radarhd} as well as material-related classification~\cite{lien2016soli,yeo2016radarcat}. Unlike active radar pipelines that rely on dedicated sensing waveforms and range-angle processing, PRISM studies material characterization from communication-oriented propagation maps. The two paradigms are complementary: radar resolves individual reflections in a sensor-centric coordinate system, whereas propagation-map sensing exploits the aggregate spatial signature produced by reflection, transmission, diffraction, and multipath across an indoor environment.}

\noindent\textcolor{blue}{\noindent{\bf Radio-Map Estimation and Digital Twins.} A related body of work reconstructs spatial radio maps from sparse measurements using interpolation~\cite{romero2022spectrum}, CNNs~\cite{yapar2023radiomapseer}, generative and Transformer-based models~\cite{zhang2023rmegan}, neural fields~\cite{zhao2023nerf2,orekondy2023winert}, or ray-tracing-assisted digital twins~\cite{hoydis2023sionnart,alkhateeb2023digitaltwin}. The primary objective of these methods is typically to predict signal power, coverage, or channel quality at unobserved locations. PRISM operates downstream of this problem: given a propagation map, it infers the location and material category of physical structures responsible for the observed propagation pattern. Therefore, radio-map estimation can serve as an upstream component of PRISM in deployments where only sparse channel measurements are available.}

{\color{blue}
\noindent{\textbf{Propagation-Map-Based Object and Material Detection.}} 
Closest to our setting, our prior RF-Vision framework~\cite{sunday2025rfv} demonstrated the feasibility of using DT-generated RF propagation maps for object characterization. However, it relied on a convolutional YOLO detector, assumed a fixed transmitter configuration, and treated material recognition as a closed-set problem. Architecturally, standard convolutional models suffer from severe spatial bottlenecks when applied to RF data, as their localized receptive fields fail to efficiently capture diffuse, multipath scattering. 

\noindent{\textbf{Motivation for \name.}} To overcome these limitations, the design of \name~introduces a fundamental paradigm shift. We identify that the ViT-DETR architecture, with its global self-attention and anchor-free bipartite matching, is mathematically uniquely suited to capture the continuous, long-range dependencies present in RF propagation maps. By training this lightweight, edge-friendly architecture entirely within a scalable DT environment, \name~preserves user privacy while extending the problem setting in three practical challenging conditions, discussed next.}

%% file: Sections/problem-formulation-new.tex
\section{System Model and Problem Formulation}
\label{sec:problem-formulation}

\subsection{Practical Challenges in RF-Based Perception}

To deploy RF-based object characterization systems in real-world MEC-enabled edge devices, the architecture must account for the physical and operational dynamics of the physical space. This work identifies three fundamental real-world challenges that are strictly necessary to address for practical deployment:

\noindent
\textbf{Challenge 1: Transmitter Mobility and Spatial Dynamics.}
In realistic IoT and 6G networked robotics environments, RF signals are not emitted from a single, static, centrally located access point. Instead, transmitters are inherently mobile (e.g., roaming robotic agents, wearable devices) or asymmetrically distributed due to physical constraints. A robust perception system must therefore extract physical material signatures that are invariant to the transmitter's spatial coordinates, ensuring reliable operation across arbitrary mobile locations.

\noindent
\textbf{Challenge 2: Environmental Domain Shift.}
Real-world indoor environments are geometrically diverse, exhibiting highly variable multipath scattering, reflection, and diffraction characteristics. A detection model deployed in a novel environment will experience a severe domain shift compared to its offline training distribution. Practical systems require a rapid adaptation mechanism to generalize to entirely unseen architectural layouts without necessitating exhaustive, large-scale data collection, and retraining at the new deployment site.

\noindent
\textbf{Challenge 3: Open-World Uncertainty.}
Autonomous systems inevitably encounter objects composed of novel materials that were not present in the offline training corpus. Traditional closed-set classification models forcefully map these out-of-distribution (OOD) materials into predefined categories, generating systematic false positives. For safety-critical and reliable autonomous navigation, it is a strict prerequisite that the perception system recognizes its own uncertainty, actively rejecting and suppressing detections for unknown materials rather than providing erroneous classifications.

\subsection{Formal Problem Formulations}
\label{subsec:problem_formulation}

Let the physical indoor environment be denoted as a scenario $\mathcal{S}$, which contains a set of physical objects $\mathcal{O}$ characterized by varying geometries and material compositions. We define the continuous location space of a mobile RF transmitter as $\mathcal{T}$. For a given scenario $\mathcal{S}$ and a specific instantaneous transmitter location $\tau \in \mathcal{T}$, the resulting spatial distribution of electromagnetic wave interference and scattering is captured as an RF propagation map, denoted as $\mathcal{M}(\mathcal{S}, \tau)$. 

The core objective of \name~is to learn an optimal detection and characterization function $f_{\theta}$, parameterized by neural network weights $\theta$, that maps the RF propagation data to the precise spatial coordinates and material properties of each object in the environment:

$$f_{\theta}: \mathcal{M}(\mathcal{S}, \tau) \rightarrow \{(b_i, c_i, s_i)\}_{i=1}^N$$

where $N$ is the number of detected objects. For the $i$-th detection:
\begin{itemize}
    \item $b_i = (x_i, y_i, w_i, h_i)$ represents the bounding box coordinates (center $x,y$, width $w$, and height $h$);
    \item $c_i \in \mathcal{Y}_{known} = \{1, 2, \dots, K\}$ represents the predicted material class label among $K$ known categories (e.g., concrete, glass, metal, wood);
    \item $s_i \in [0, 1]$ represents the calibrated confidence score associated with the prediction.
\end{itemize}

Based on this general system model, we formalize the three specific problems addressed in this paper:

\noindent
\textbf{Problem Formulation 1 (Spatial Invariance under Transmitter Mobility):}
Let $p(\tau)$ denote the probability distribution of the mobile transmitter's location across the trajectory space $\mathcal{T}$. If a model is trained exclusively on a static location $\tau_0$, the learned function implicitly conditions on that spatial origin, $f_\theta(\cdot \mid \tau_0)$, leading to a high generalization error when $\tau \sim p(\tau) \neq \tau_0$. The objective is to design a feature extraction architecture that learns a \emph{location-invariant} mapping, ensuring that the detection loss $\mathcal{L}$ remains stable regardless of the transmitter's location:
$$\mathbb{E}_{\tau \sim p(\tau)} \left[ \mathcal{L}\left(f_\theta(\mathcal{M}(\mathcal{S}, \tau)), Y\right) \right] \approx \min_{\theta} \mathcal{L}\left(f_\theta(\mathcal{M}(\mathcal{S}, \tau_0)), Y\right)$$
where $Y$ is the ground-truth set of objects. This ensures consistent localization and characterization across the full mobility space $\mathcal{T}$.

\noindent
\textbf{Problem Formulation 2 (Cross-Scenario Few-Shot Adaptation):}
Let $\mathcal{D}_{source} = \{(\mathcal{M}(\mathcal{S}_{train}, \tau), Y)\}$ represent the data distribution of the training scenarios, and $\mathcal{D}_{target} = \{(\mathcal{M}(\mathcal{S}_{new}, \tau), Y)\}$ represent an entirely unseen deployment environment, where $\mathcal{S}_{train} \cap \mathcal{S}_{new} = \emptyset$. Due to structural multipath variations, the marginal distributions of the RF maps differ significantly: $P_{source}(\mathcal{M}) \neq P_{target}(\mathcal{M})$. The objective is to maximize the expected detection accuracy on $\mathcal{D}_{target}$ by updating $f_\theta$ using only a highly constrained support set $S \subset \mathcal{D}_{target}$ (where $|S|$ is strictly minimal), thereby achieving rapid few-shot domain adaptation.

\noindent
\textbf{Problem Formulation 3 (Open-Set Anomaly Rejection):}
In an open-world deployment, the environment contains objects whose true material class $y_i$ belongs to an unknown set $\mathcal{Y}_{unknown}$, such that $\mathcal{Y}_{known} \cap \mathcal{Y}_{unknown} = \emptyset$. The goal is not to explicitly classify these anomalous materials, but rather to formulate a robust rejection mechanism. The model must produce no bounding box output for out-of-distribution materials to prevent systematic false positives. Formally, the desired inference behavior necessitates a gating mechanism based on a calibrated confidence threshold $\delta$:
$$f_\theta\bigl(\mathcal{M}(\mathcal{S}, \tau)\bigr) = \{(b_i, c_i, s_i)\}_{i=1}^{N'}$$
where $N' \leq N$. The output set contains \emph{only} detections where $c_i \in \mathcal{Y}_{known}$ and $s_i \geq \delta$. All spatial features activated by $y_i \in \mathcal{Y}_{unknown}$ are actively suppressed, yielding an empty detection output for those specific physical regions.

%% file: Sections/proposed-framework-new.tex
\section{$\name$ Framework}
\label{sec:proposed_framework}
\vspace*{-3pt}


In this section, we present different components of proposed framework solving the problems formulated in the Section~\ref{subsec:problem_formulation}.

\begin{figure*}[t!]
    \centering
    \includegraphics[width=1\linewidth]{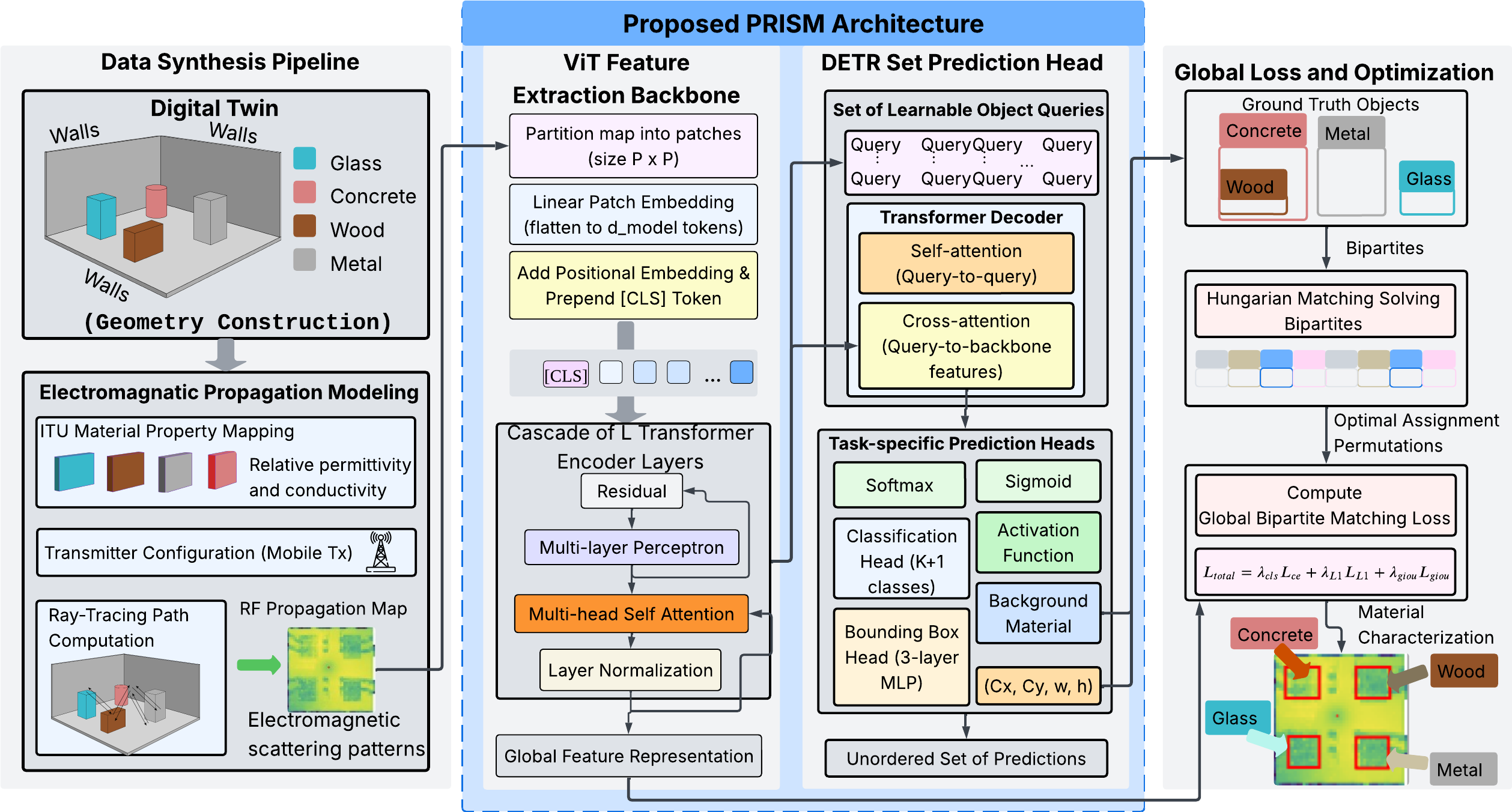}
    \caption{The proposed \name~detection architecture, integrating a feature extraction backbone with a set prediction head to achieve location-invariant material characterization. The detailed data synthesis pipeline is further described in Section~\ref{sec:testbed_design}.}
    \label{fig:framework_arch}
\end{figure*}

\label{subsec:framework-transmitter-mobility}

\subsection{Transmitter Mobility-Agnostic Detection via ViT-DETR}
\label{subsec:framework-transmitter-mobility}

To solve the spatial invariance challenge defined in first problem in Section~\ref{subsec:problem_formulation}, we must design a function $f_\theta$ capable of maintaining a stable detection loss regardless of the mobile transmitter's instantaneous location $\tau$. Traditional anchor-based convolutional architectures are fundamentally ill-suited for this task, as they rely on localized receptive fields that overfit to specific spatial interference artifacts. 

{\color{blue}
To overcome this, we propose \name, a lightweight, end-to-end material detector that formulates object characterization as a direct set-prediction problem. While ViT and DETR components are established in computer vision, their application in \name~is motivated by RF physics. Instead of localized receptive fields, the ViT backbone utilizes global self-attention to capture diffuse scattering and long-range electromagnetic dependencies, structurally mirroring how a propagating wave interacts with multiple surfaces simultaneously. Furthermore, rather than relying on rigid spatial anchors, our DETR-style head uses dynamic object queries. This anchor-free approach allows the model to localize objects based on their holistic scattering footprint, directly enabling the location-invariant feature learning required for mobile transmitter robustness. As illustrated in the comprehensive block diagram in Fig.~\ref{fig:framework_arch}, the framework is structurally divided into a Data Synthesis Pipeline feeding into a two-stage machine learning architecture optimized to learn the location-invariant mapping $\mathbb{E}_{\tau \sim p(\tau)} \left[ \mathcal{L}\left(f_\theta(\mathcal{M}(\mathcal{S}, \tau)), Y\right) \right]$.}


\subsubsection{Feature Extraction Backbone}
As shown in Block II of Fig.~\ref{fig:framework_arch}, the network ingests the RF propagation map $\mathcal{M}(\mathcal{S}, \tau) \in \mathbb{R}^{C \times H \times W}$ synthesized from a specific transmitter configuration $\tau$. The map is first partitioned into $N$ non-overlapping patches of size $P \times P$. These patches undergo linear patch embedding to flatten them into $d_{model}$-dimensional tokens. 

To retain the geometric configuration of the original RF scenario, a learnable 1D position embedding is added to the patch sequence, which is then prepended with a dedicated \texttt{[CLS]} token. The resulting sequence of $N_{tokens} = (H/P \times W/P) + 1$ tokens is processed through a cascade of $L$ Transformer encoder layers. Each layer consists of multi-head self-attention (MSA) and a multi-layer perceptron (MLP) block connected via residual connections:

\begin{equation}
    \mathbf{z}_{\ell}' = \text{MSA}(\text{LN}(\mathbf{z}_{\ell-1})) + \mathbf{z}_{\ell-1}, \quad \ell=1, \dots, L
\end{equation}
\begin{equation}
    \mathbf{z}_{\ell} = \text{MLP}(\text{LN}(\mathbf{z}_{\ell}')) + \mathbf{z}_{\ell}'
\end{equation}
where $\text{LN}$ denotes Layer Normalization. The final encoder output $\mathbf{z}_{L} \in \mathbb{R}^{B \times N_{tokens} \times d_{model}}$ serves as a rich, globally aware electromagnetic feature representation. {\color{blue} By correlating signals across the entire map, the MSA layers build a
global representation of the illumination geometry, including the
transmitter position and the dominant scatterers. We emphasize that the
encoder does not discard $\tau$: as shown in Section~\ref{sec:attn},
it attends strongly to the transmitter location. Transmitter invariance
instead emerges at the decoder, whose object queries read out
material-bearing regions with attention that is nearly uniform over the
objects and insensitive to where the transmitter lies.}

\subsubsection{Set Prediction Head}
The extracted features are passed to the prediction head, which eliminates the need for hand-crafted anchor generation by employing a Transformer decoder. We define a fixed set of $N_q$ learnable object queries.

Within the Transformer decoder, these object queries first undergo self-attention (query-to-query) to model inter-object relationships, preventing multiple queries from collapsing onto the same highly reflective object. This is followed by cross-attention (query-to-backbone features), where the queries interact with the global features $\mathbf{z}_{L}$ to extract localized material signatures. The updated query embeddings are subsequently fed into two parallel, task-specific prediction heads:

\begin{itemize}
    \item \textbf{Classification Head:} A linear projection layer coupled with a \texttt{Softmax} activation function outputs probabilities for $(K + 1)$ classes. $K$ represents the total number of known materials (e.g., Concrete, Glass, Metal, Wood), while the $+1$ designates the \texttt{Background Material} category, laying the foundation for our open-set detection objective. 
    \item \textbf{Bounding Box Regression Head:} A 3-layer MLP equipped with a terminal \texttt{Sigmoid} activation function regresses the normalized continuous spatial coordinates $(c_x, c_y, w, h)$, representing the bounding box center, width, and height.
\end{itemize}

\subsubsection{Global Loss and Optimization}
To optimize the network parameters $\theta$, the architecture operates on the unordered set of $N_q$ predictions (Block IV). To supervise these predictions using a ground truth set of $M$ objects $Y$ (where $M \le N_q$), we perform global bipartite matching via the Hungarian algorithm. The algorithm identifies the optimal assignment permutation $\hat{\sigma}$ that minimizes the matching cost $\mathcal{C}_{match}$:

\begin{equation}
    \mathcal{C}_{match} = \lambda_{cls}\mathcal{C}_{cls} + \lambda_{L1}\mathcal{C}_{L1} + \lambda_{giou}\mathcal{C}_{giou}
    \label{eq:c-match}
\end{equation}
where $\mathcal{C}_{cls}$ represents the negative predicted probability of the ground truth class, $\mathcal{C}_{L1}$ is the $L_1$ spatial distance, and $\mathcal{C}_{giou}$ is the Generalized Intersection over Union (GIoU). 

Once the optimal assignments are determined, the network computes the global bipartite matching loss $\mathcal{L}_{total}$:
\begin{equation}
    \mathcal{L}_{total} = \lambda_{cls} \mathcal{L}_{ce} + \lambda_{L1} \mathcal{L}_{L1} + \lambda_{giou} \mathcal{L}_{giou}
    \label{eq:l-total}
\end{equation}
where $\mathcal{L}_{ce}$ is the cross-entropy classification loss. 

By continuously backpropagating this total loss over a broad joint distribution of physical propagation geometries (sampled from $\tau \sim p(\tau)$), the network weights $\theta$ are iteratively updated.

\begin{figure*}[t!]
    \centering
    \includegraphics[width=1\textwidth]{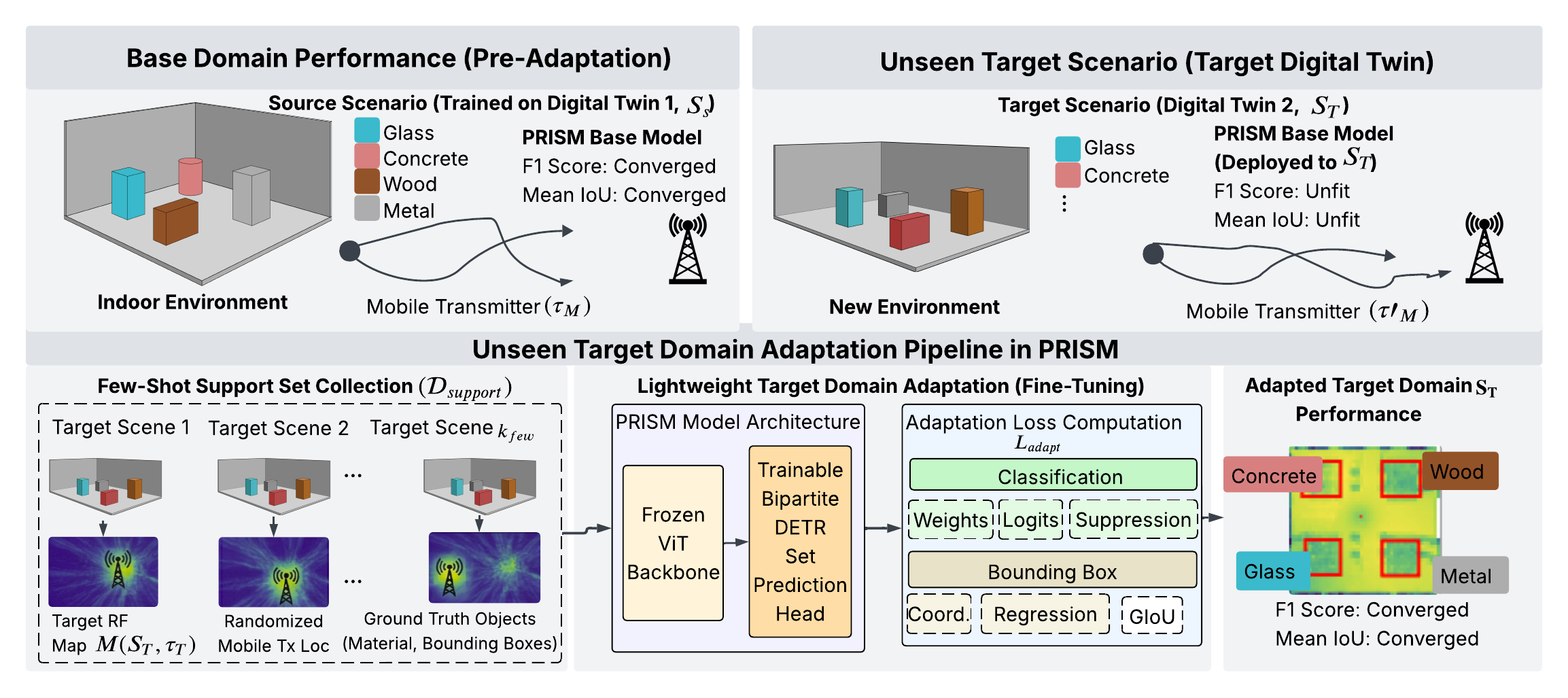}
    \caption{The \name~few-shot support set adaptation pipeline. The architecture overcomes the baseline performance drop in an unseen target scenario (top) by fine-tuning the prediction head on a minimal support set of $k$ samples, rapidly restoring high-fidelity material characterization.}
    \label{fig:solution2}
\end{figure*}

\subsection{Few-Shot Support Set Adaptation for Unseen Scenarios}
\label{subsec:framework-few-shot}

To rigorously evaluate the generalization capability of \name~and address the cross-scenario domain shift defined in the second problem of Section~\ref{subsec:problem_formulation}, we formalize a few-shot domain adaptation strategy. When deployed in a novel physical environment $\mathcal{S}_{new}$, the model confronts a domain shift driven by unseen architectural layouts and multipath configurations. This shift implies that the marginal distribution of the target RF maps diverges from the source training data, i.e., $P_{source}(\mathcal{M}) \neq P_{target}(\mathcal{M})$. To bridge this gap and maximize expected detection accuracy on $\mathcal{D}_{target}$ without full retraining, we employ a support set fine-tuning paradigm.

Let the baseline model, pre-trained on the source distribution $\mathcal{D}_{source}$, be denoted as $f_{\theta_{source}}$. We logically decompose the network parameters into two distinct sets: $\theta = \{\psi, \phi\}$, where $\psi$ represents the weights of the ViT feature extraction backbone, and $\phi$ represents the weights of the DETR-style set-prediction head. Thus, the forward pass can be expressed as $f_\theta(\cdot) = h_\phi(g_\psi(\cdot))$.

To adapt the model to the target domain, we sample a highly constrained support set $\mathcal{D}_{support} \subset \mathcal{D}_{target}$, consisting of only $k$ labeled examples (i.e., $|\mathcal{D}_{support}| = k \ll |\mathcal{D}_{target}|$). This strategy mimics realistic deployment constraints in MEC networks, where acquiring extensive empirical RF data at a newly deployed site is practically infeasible.

During the lightweight fine-tuning phase, the highly parameterized ViT backbone remains entirely frozen ($\psi \leftarrow \psi_{source}$). This prevents catastrophic overfitting on the minimal support set $\mathcal{D}_{support}$ and preserves the fundamental, domain-agnostic electromagnetic feature extraction capabilities learned during baseline training. Conversely, the DETR set-prediction head $\phi$ is actively updated via the bipartite matching adaptation loss. Formally, this targeted calibration is achieved by finding the optimal head parameters $\phi^*$ that minimize the expected loss strictly over the support set:

\begin{equation}
    \phi^* = \arg\min_{\phi} \mathbb{E}_{(\mathcal{M}, Y) \in \mathcal{D}_{support}} \left[ \mathcal{L}_{total}\left(h_\phi(g_{\psi_{source}}(\mathcal{M})), Y\right) \right]
\end{equation}

As illustrated in Fig.~\ref{fig:solution2}, this domain adaptation pipeline systematically addresses the performance degradation encountered when the baseline model is deployed in an entirely unseen target scenario (e.g., transitioning from the structured \textit{Standard Split} to the \textit{Cross-Scenario Split} as evaluated in Section~\ref{subsec:exp2_few_shot}). By updating the network weights via $S$, this targeted calibration rapidly realigns the model to the specific multipath distribution $P_{target}(\mathcal{M})$ of the new domain. This approach successfully fully restores spatial localization and material classification fidelity while bootstrapping generalization with minimal data acquisition overhead.



\begin{figure}[t!]
    \centering
\includegraphics[width=\linewidth]{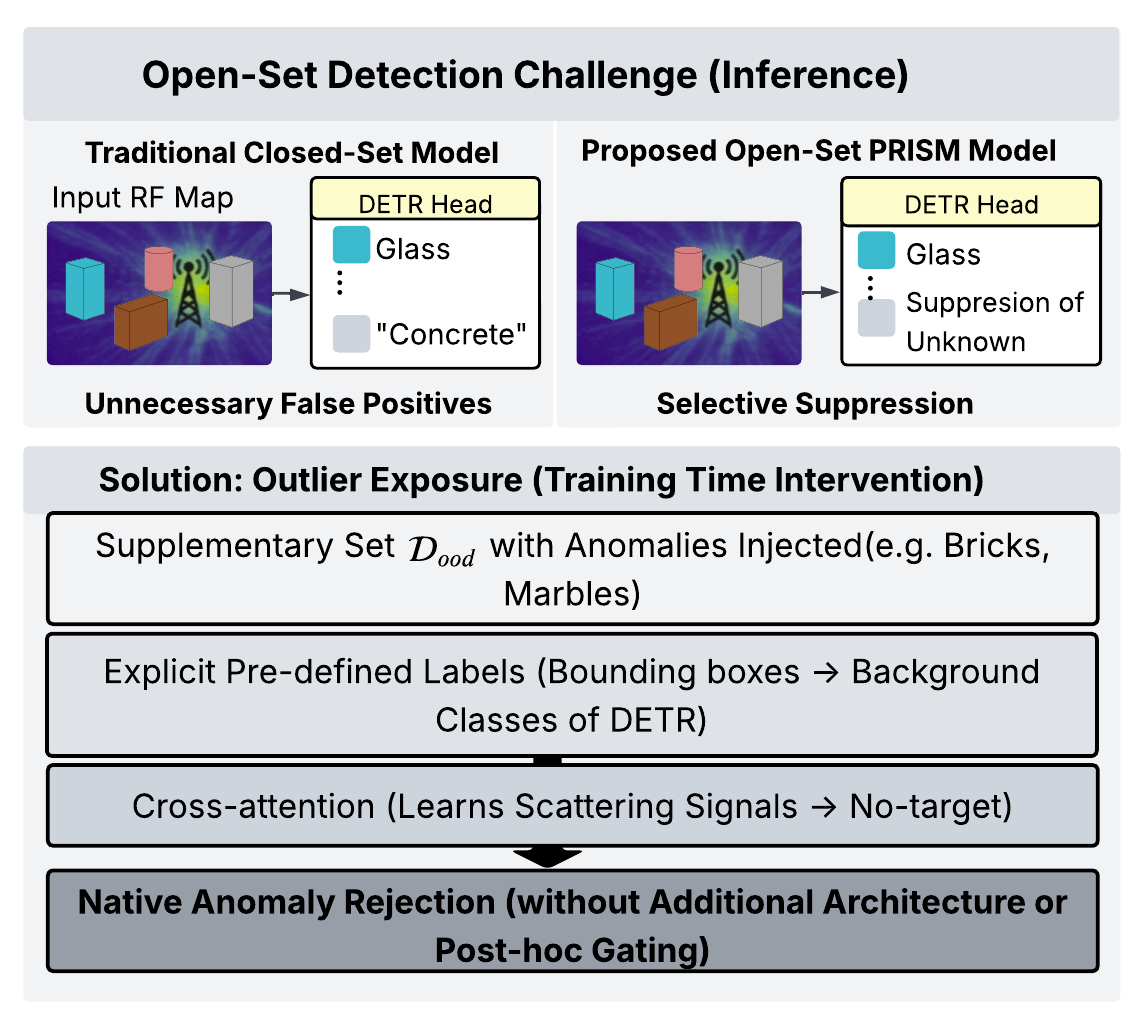}
    \caption{The \name~open-set unknown material rejection pipeline via outlier exposure. By explicitly matching OOD materials to the DETR \texttt{background} class during fine-tuning, the architecture learns to actively suppress anomalous detections at inference rather than forcing erroneous known-class assignments.}
    \label{fig:solution3}
\end{figure}

\subsection{Open-Set Unknown Material Rejection via Outlier Exposure}
\label{subsec:framework-open-set}

To address the open-world deployment constraints defined in the third problem of Section~\ref{subsec:problem_formulation}, the detector is targeted to actively suppress predictions on novel materials $y_i \in \mathcal{Y}_{unknown}$ without modifying the primary known-class detection architecture. To achieve this, we integrate an outlier exposure strategy, as depicted in Fig.~\ref{fig:solution3}. 

In dynamic settings, the model must maintain detection fidelity for $\mathcal{Y}_{known}$ while simultaneously rejecting OOD materials. The core principle of outlier exposure is to present a limited number of OOD material samples to the model during the fine-tuning phase, explicitly constraining them to the \texttt{background} class label rather than forcing a geometric assignment to a known material. 

We use the $\mathcal{D}_{support}$ from Section~\ref{subsec:framework-few-shot} as known-class ($\mathcal{Y}_{known}$) support set, and denote $\mathcal{D}_{ood}$ as a small supplementary set of outlier samples representing the unknown material classes ($\mathcal{Y}_{unknown}$). The adaptation loss objective is applied over the union dataset $\mathcal{D}_{support} \cup \mathcal{D}_{ood}$. During the global bipartite matching step of the DETR head, the ground-truth labels for the spatial coordinates of samples in $\mathcal{D}_{ood}$ are strictly mapped to the \texttt{background} classification target. This formulation structurally incentivizes the cross-attention layers to associate anomalous electromagnetic scattering signatures with the absolute absence of a valid target. 

By forcing OOD features into the background manifold during fine-tuning, we natively realize the inference gating mechanism postulated in the third formulated problem in Section~\ref{subsec:problem_formulation}. During inference on an unconstrained RF map $\mathcal{M}(\mathcal{S}, \tau)$, spatial regions exhibiting physical properties dissimilar to the trained materials trigger high background probabilities. Consequently, the predicted confidence scores $s_i$ for any known class $c_i \in \mathcal{Y}_{known}$ fall well below the calibrated detection threshold $\delta$. This gating mechanism guarantees that all spatial queries activated by $y_i \in \mathcal{Y}_{unknown}$ are effectively filtered out, yielding the desired strict subset of reliable outputs $N' \leq N$ and successfully preventing systematic false positives in open-world MEC scenarios.



%% file: Sections/dataset-new.tex
\section{Testbed Design and Data Synthesis}
\label{sec:testbed_design}

\begin{figure*}[t!]
    \centering    
    \includegraphics[width=\linewidth]{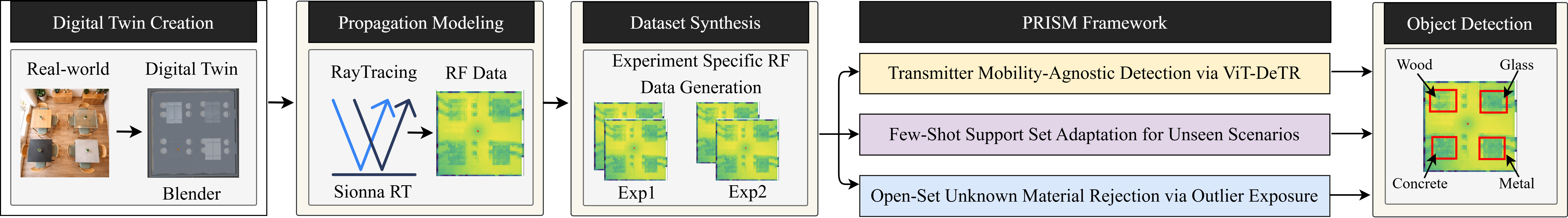}
    \caption{The flow diagram of \name~testbed.}
    \label{fig:system_overview}
    \vspace{-10pt}
\end{figure*}

In this section, we detail the construction of the \name~dataset, a comprehensive, ray-tracing-based repository designed for robust object characterization in indoor environments. By employing deterministically modeled DTs, this dataset provides a high degree of physical realism while guaranteeing accurate ground-truth annotations. We outline the end-to-end data synthesis pipeline: from geometric modeling to electromagnetic propagation simulation to facilitate reproducibility. The overall flow of the pipeline is shown in Fig.~\ref{fig:system_overview}. The complete codebase, digital twin models, and ray-tracing datasets are publicly available \cite{PRISM2024}.

\noindent
\textbf{Experimental Platform.} The \name~dataset is synthesized using a high-performance workstation (Alienware 18 Area-51, Ubuntu OS). The geometric environment construction is performed using Blender v4.2.17 LTS \cite{blender}, while the electromagnetic wave propagation and ray-tracing are executed via NVIDIA's Sionna RT framework \cite{hoydis2023sionna}. 

\subsection{Environmental Configurations}
\label{Sec:scenario}
To evaluate the spatial robustness of the proposed framework, we construct two distinct architectural scenarios inspired by real-world academic environments: a semi-structured laboratory cubicle space and an open-plan meeting area. The spatial coordinates and material properties (e.g., wood, metal, glass, concrete) of the objects within these scenes are dynamically varied every 500 samples to ensure distributional diversity.

\begin{figure*}[t!]
	\centering
	\begin{subfigure}{0.24\linewidth}
        \centering        
        \includegraphics[width=\linewidth]{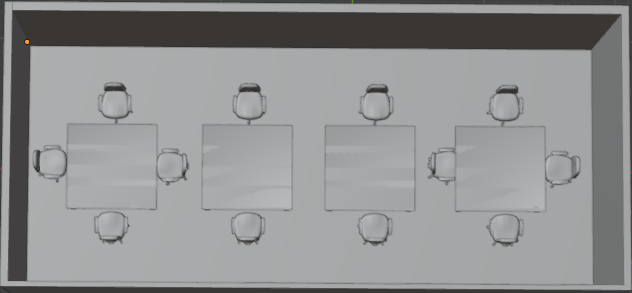}
		\caption{Cubicle\_Area (DT)}
	\end{subfigure}
	\begin{subfigure}{0.24\linewidth}
        \centering
        \includegraphics[width=\linewidth]{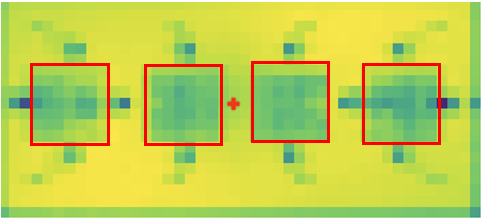}
        \caption{Cubicle RF Map w/ TX @[-3.3, -0.5]}
	\end{subfigure}
    \begin{subfigure}{0.24\linewidth}
        \centering        
        \includegraphics[width=\linewidth]{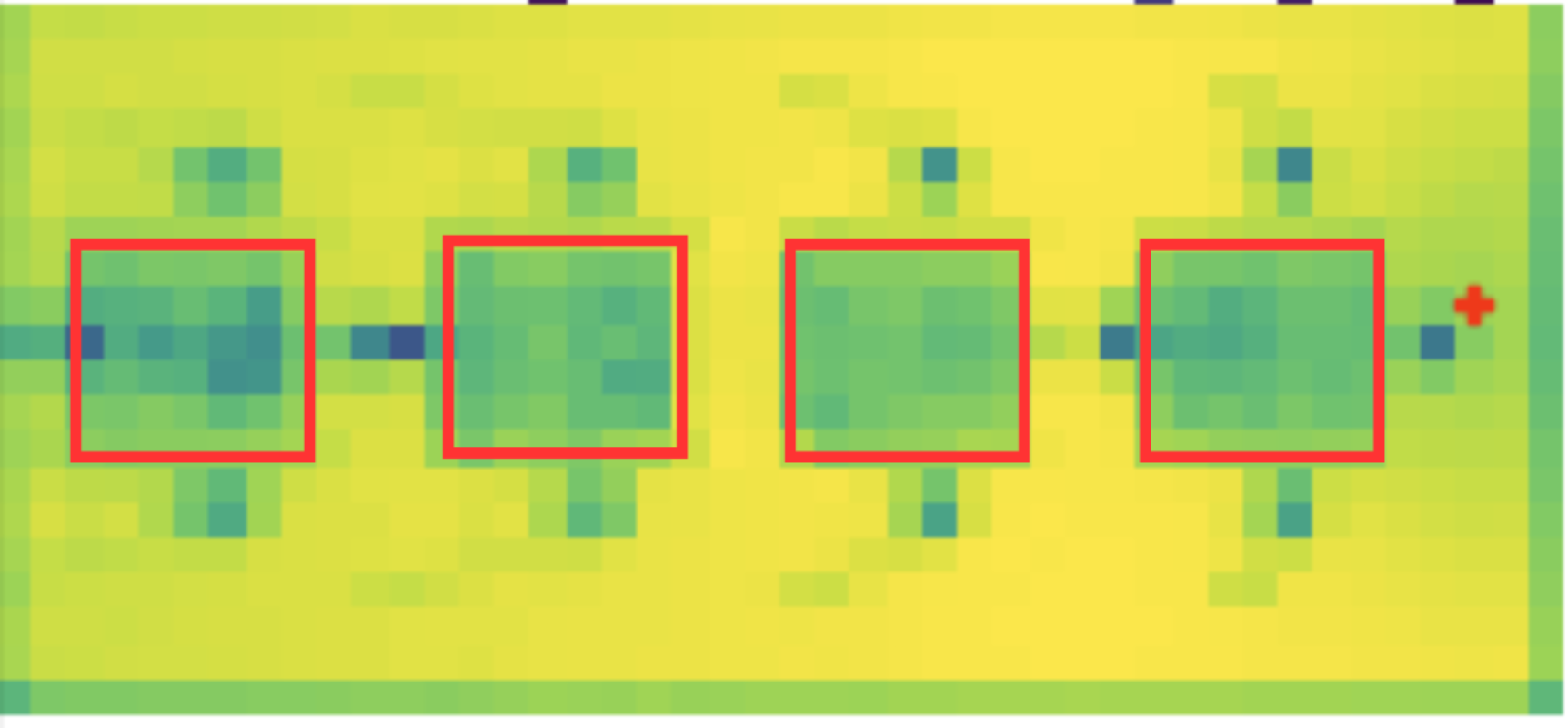}
		\caption{Cubicle RF Map w/ TX @[-23, -2.4]}
	\end{subfigure}
	\begin{subfigure}{0.24\linewidth}
        \centering
        \includegraphics[width=\linewidth]{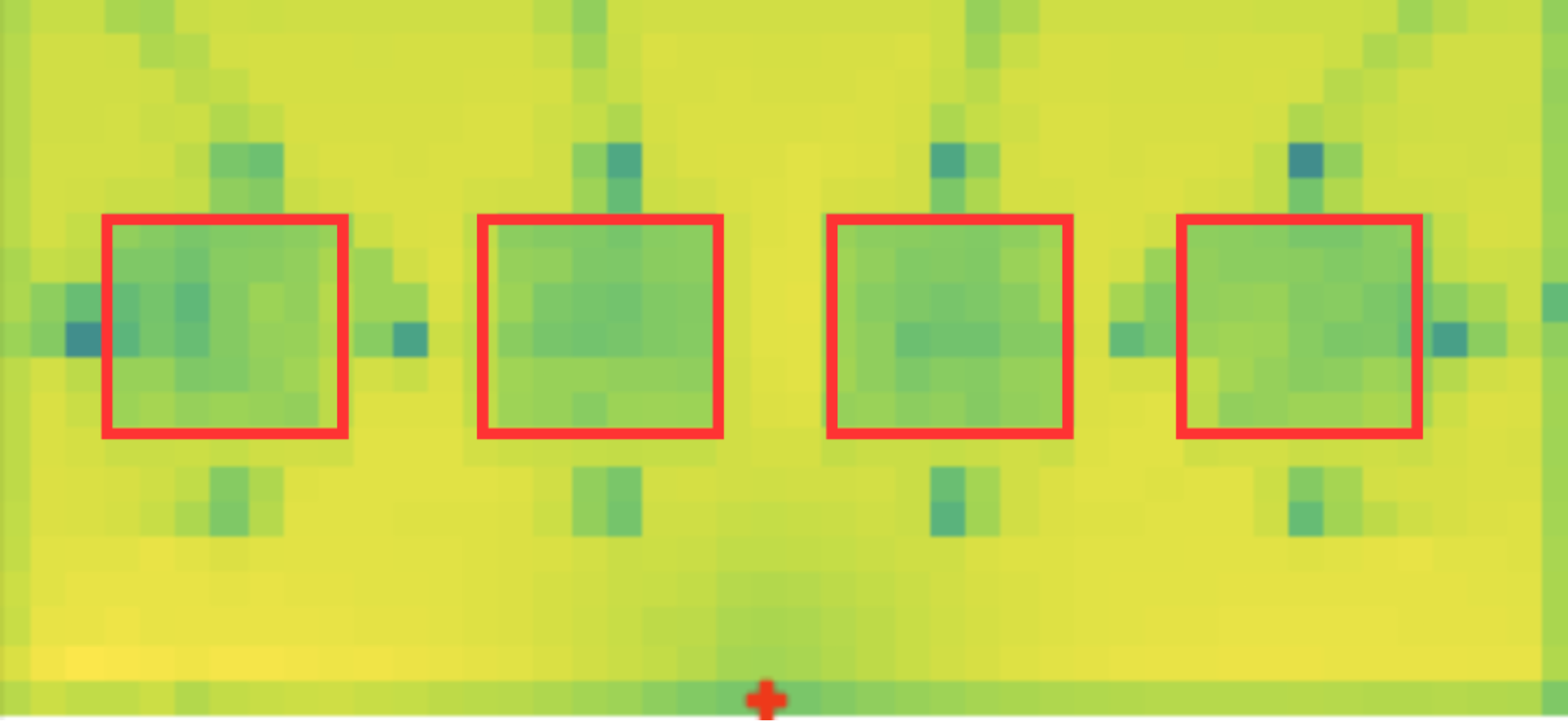}
        \caption{Cubicle RF Map w/ TX @[-3.5, 9.2]}
	\end{subfigure}

	\begin{subfigure}{0.24\linewidth}
        \centering
        \includegraphics[width=\linewidth]{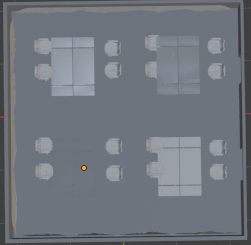}
		\caption{Meeting\_Area 1 (DT)}
	\end{subfigure}
	\begin{subfigure}{0.24\linewidth}
        \centering
        \includegraphics[width=\linewidth]{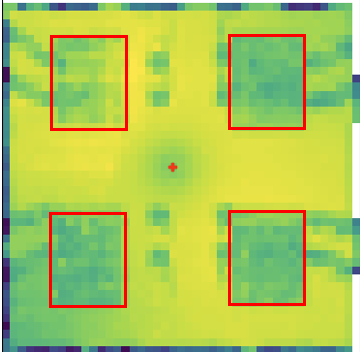}
        \caption{Meeting 1 RF Map w/ TX @[1.4, 0.9]}
	\end{subfigure}
    \begin{subfigure}{0.24\linewidth}
        \centering
        \includegraphics[width=\linewidth]{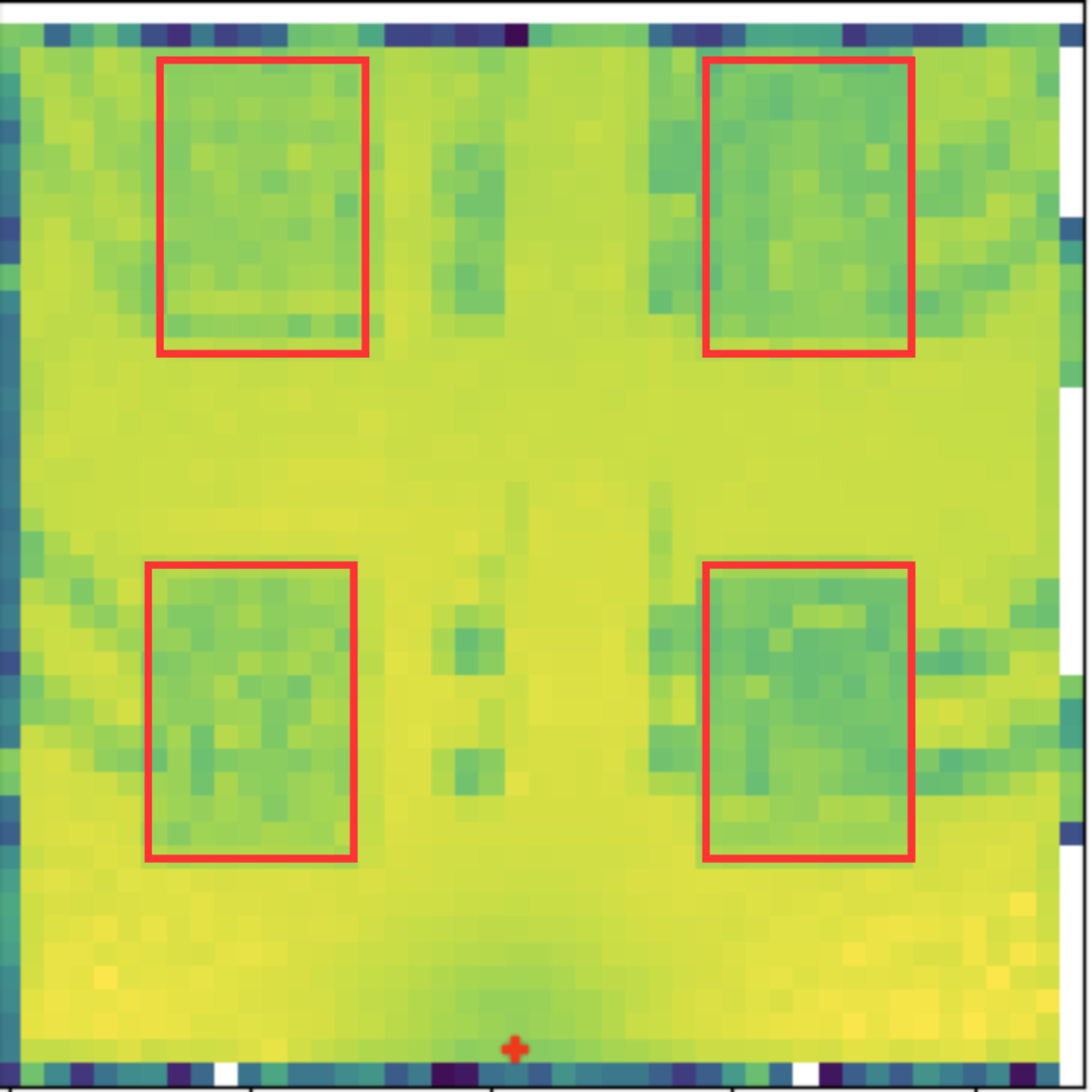}
		\caption{Meeting 1 RF Map w/ TX @[0.8, -21]}
	\end{subfigure}
	\begin{subfigure}{0.24\linewidth}
        \centering
        \includegraphics[width=\linewidth]{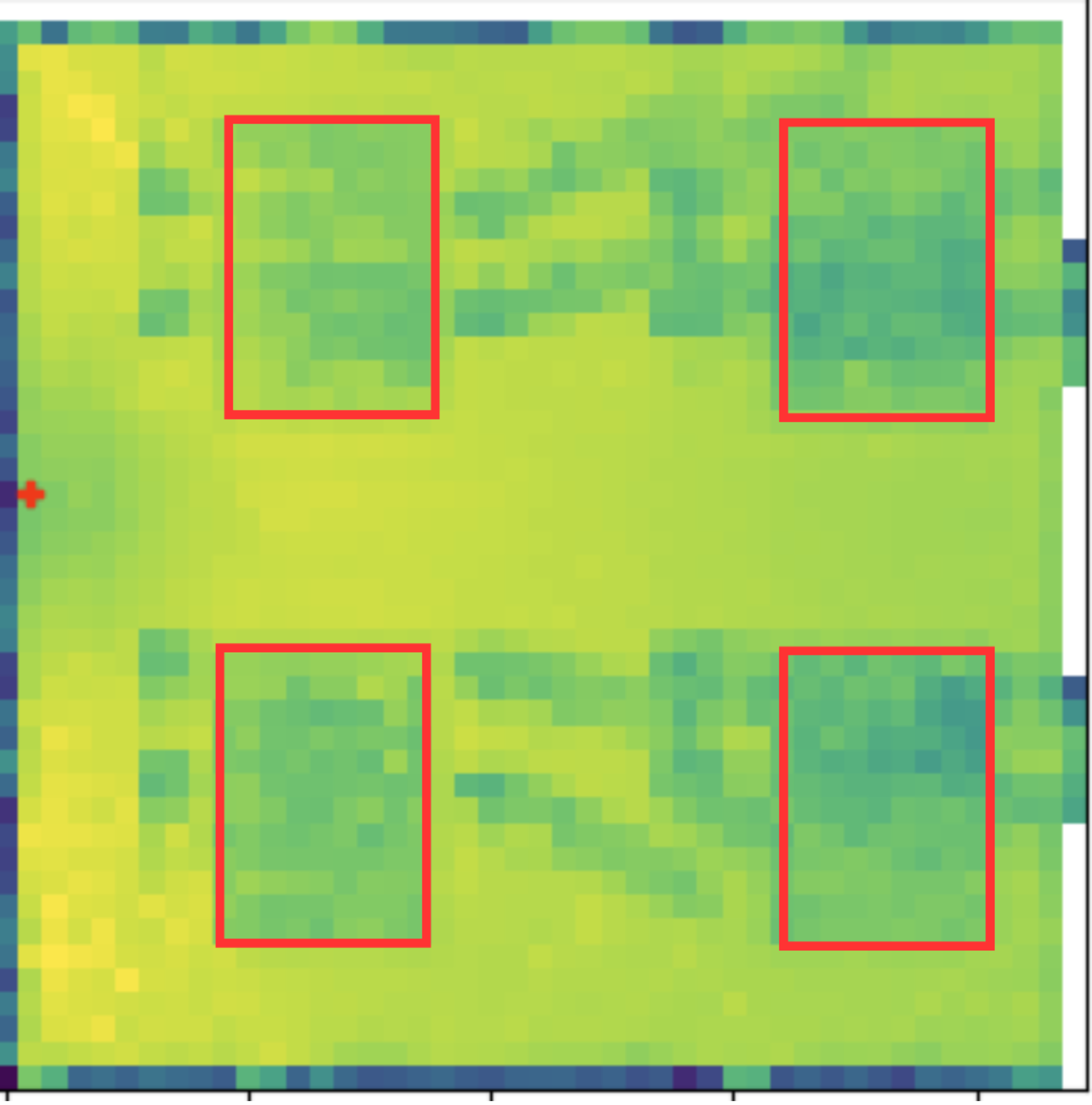}
        \caption{Meeting 1 RF Map w/ TX @[-19, -0.7]}
	\end{subfigure} 

	\begin{subfigure}{0.24\linewidth}
        \centering
        \includegraphics[width=\linewidth]{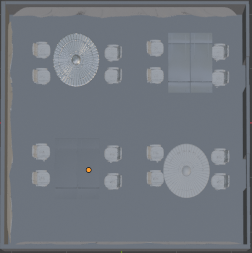}
		\caption{Meeting\_Area 2 (DT)}
	\end{subfigure}
	\begin{subfigure}{0.24\linewidth}
        \centering
        \includegraphics[width=\linewidth]{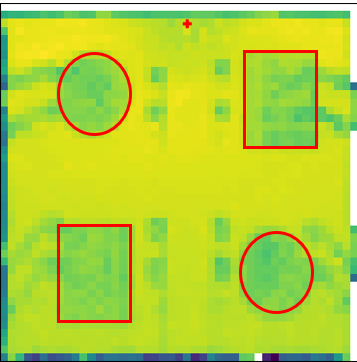}
        \caption{Meeting 2 RF Map w/ TX @[1.3, 19]}
	\end{subfigure}
    \begin{subfigure}{0.24\linewidth}
        \centering
        \includegraphics[width=\linewidth]{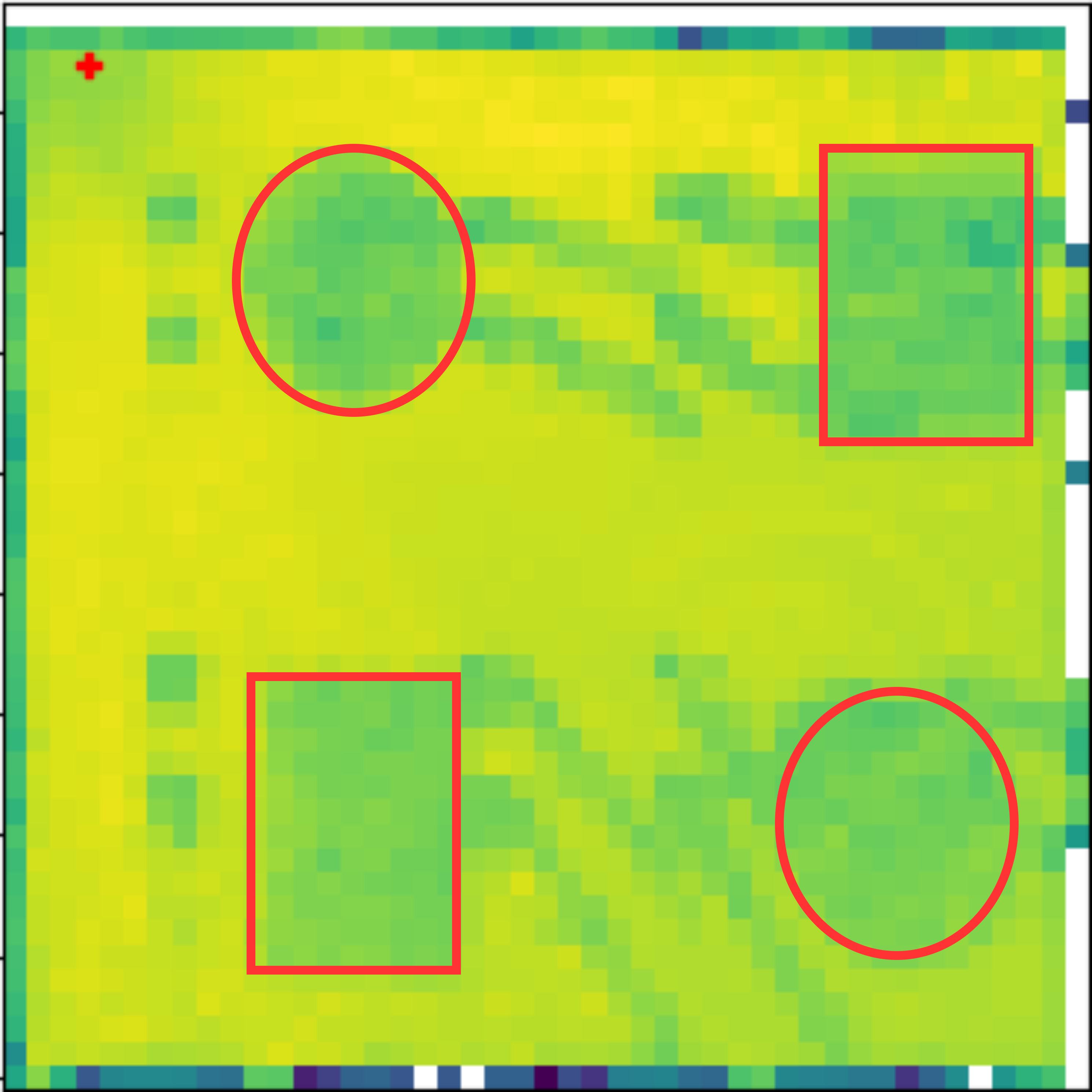}
		\caption{Meeting 2 RF Map w/ TX @[-18, 18]}
	\end{subfigure}
	\begin{subfigure}{0.24\linewidth}
        \centering
        \includegraphics[width=\linewidth]{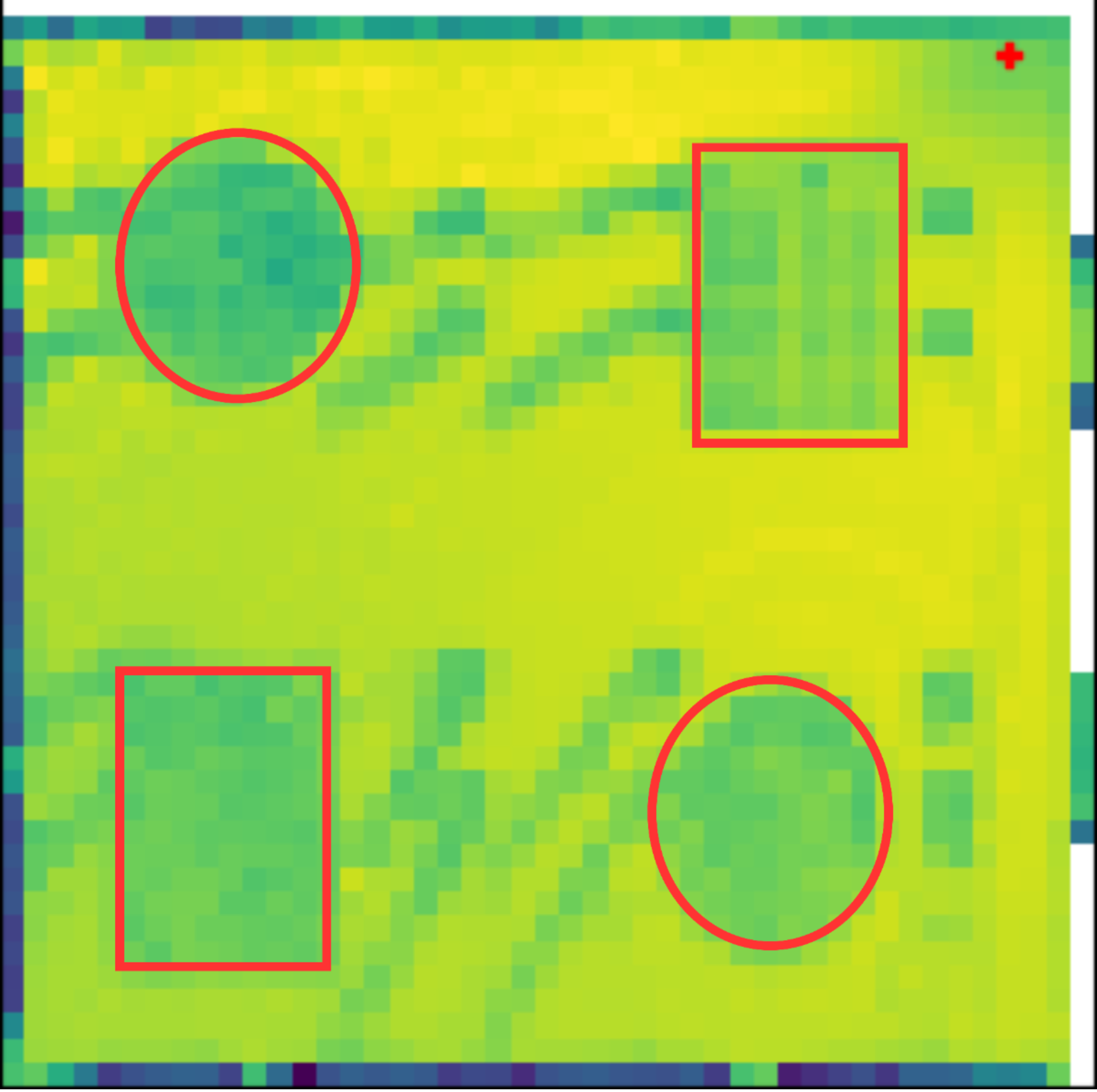}
        \caption{Meeting 2 RF Map w/ TX @[18, 19]}
	\end{subfigure}
    	\begin{subfigure}{0.24\linewidth}
        \centering
        \includegraphics[height=4.5cm] {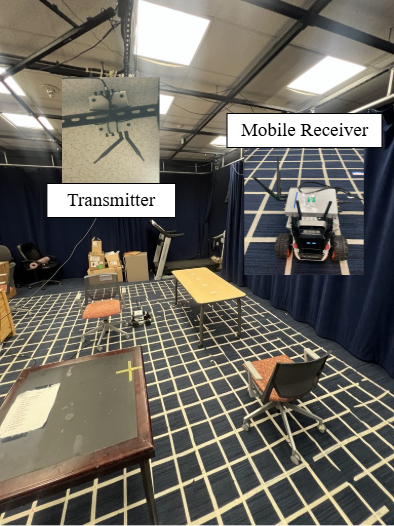}
		\caption{\textcolor{blue}{Real world Testbed}}
	\end{subfigure}
	\begin{subfigure}{0.24\linewidth}
        \centering
        \includegraphics[width=\linewidth]{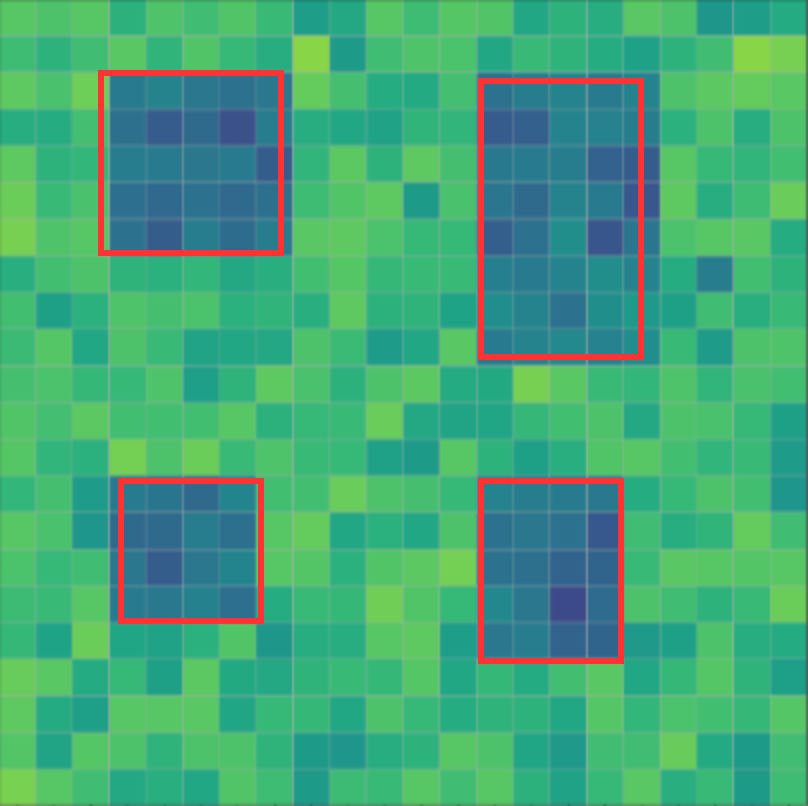}
        \caption{\textcolor{blue}{Real world RF Sample 1}}
	\end{subfigure}
    \begin{subfigure}{0.24\linewidth}
        \centering
        \includegraphics[width=\linewidth]{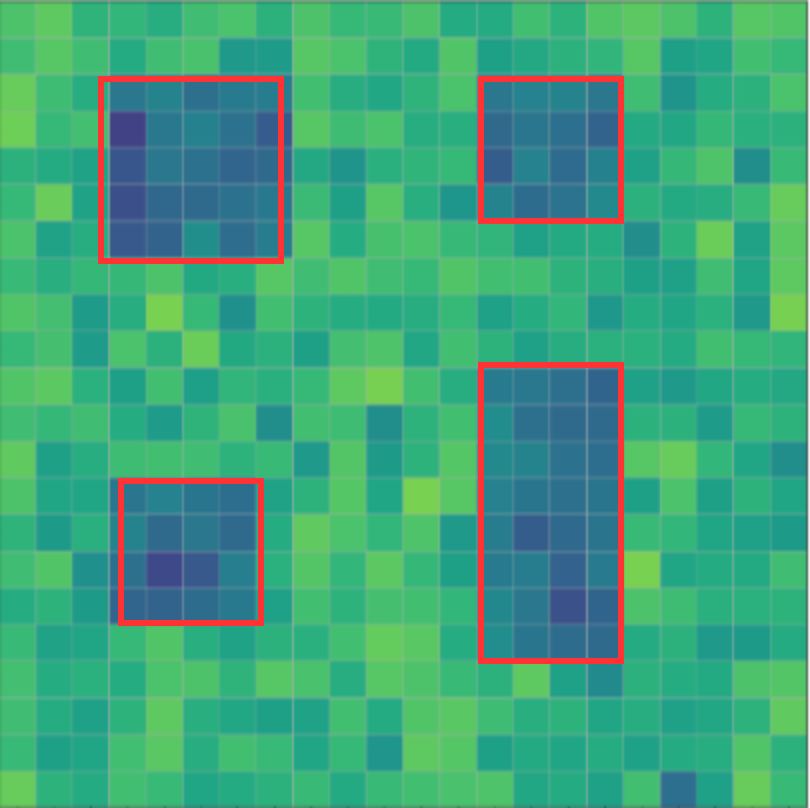}
		\caption{\textcolor{blue}{Real world RF Sample 2}}
	\end{subfigure}
	\begin{subfigure}{0.24\linewidth}
        \centering
        \includegraphics[width=\linewidth]{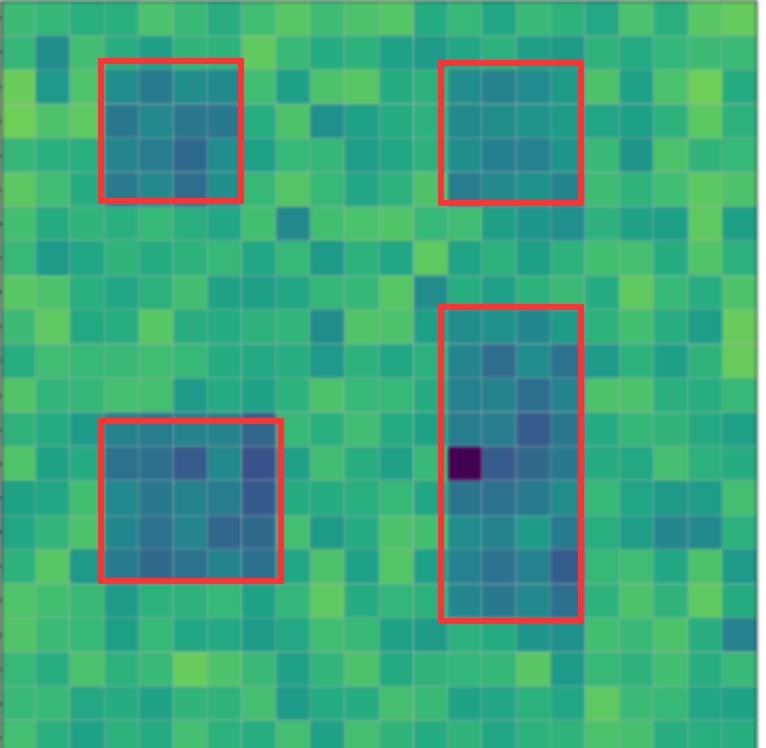}
        \caption{\textcolor{blue}{Real world RF Sample 3}}
	\end{subfigure}
  \caption{\textcolor{blue}{A snapshot of PRISM dataset. Panels (a), (e), and (i) display the geometric DTs constructed in Blender, while panel (m) shows the real-world testbed. The adjacent images represent the corresponding RF propagation maps synthesized using Sionna RT and the RF samples collected from the real-world testbed. Highlighted regions indicate object positions and areas of high-intensity electromagnetic scattering, while the physical location of the mobile transmitter is designated by a red cross (+) in the DT scenes and by ``TX'' in the real-world testbed.}}
    \vspace{-0.2in}
	\label{fig:Dataset}
\end{figure*}

\begin{itemize}
    \item \textbf{Cubicle\_Area:} This environment represents a highly constrained, multipath-rich layout measuring $42 \times 18$ meters (bounded by coordinates $[-24, -9]$ and $[18,9]$). It contains 4 tables, 12 chairs, and narrow corridors, fully enclosed by surrounding walls. This setting induces severe 
    NLoS and complex diffraction patterns. The geometric layout is illustrated in Fig.~\ref{fig:Dataset}(a).
    
    \item \textbf{Meeting\_Area:} Representing a more expansive, open-plan environment, this scenario measures $42 \times 41$ meters (bounded by coordinates $[-20,-22]$ and $[22,19]$). It contains 4 tables and 16 clustered chairs, enclosed by bounding walls. This scenario evaluates the model's performance in environments dominated by line-of-sight (LoS) and specular ground reflections. The layout is illustrated in Fig.~\ref{fig:Dataset}(e) and (i).
\end{itemize}

{\color{blue}
\subsection{Real-World RF Data Collection Campaign}
\label{subsec:real_world_campaign}

To systematically bridge the sim-to-real gap and evaluate the robustness of the \name~framework under real-world RF propagation conditions, we conducted an empirical measurement campaign using a physical Software Defined Radio (SDR) testbed. The experimental setup and data collection procedure are described below.

\noindent$\bullet$ \textbf{Physical layout:}
The experiments were conducted in a $22 \times 22$ foot indoor physical grid containing four obstacles, comprising two wooden tables and two plastic chairs, positioned near the center of the measurement area. These obstacles introduce NLoS propagation conditions, signal attenuation, and multipath reflections, creating a challenging RF environment with spatial coverage variations.

\noindent$\bullet$ \textbf{Hardware and data collection:}
The hardware setup consists of two USRP B210 SDRs, with one configured as a mobile legitimate transmitter (Tx) transmitting a continuous modulated signal and the other serving as a mobile receiver (Rx). Both SDRs operate at a center frequency of $915$ MHz. The transmitter is configured with a sampling rate of $1$ MS/s and a gain setting of $85$ dB, while the receiver operates at a sampling rate of $20$ MS/s with a gain setting of $76$ dB. At each measurement location, the receiver utilizes the UHD FFT-based spectrum visualization tool (\texttt{uhd\_fft}) to record the peak Received Signal Strength (RSS) value of the transmitted signal. For each measurement snapshot, the receiver was systematically traversed across every predefined grid point within the $22 \times 22$ foot indoor environment to collect RSS measurements, thereby generating a complete spatial RSS map. To introduce environmental diversity, the four physical obstacles were randomly repositioned after each snapshot, creating distinct RF propagation conditions. This measurement campaign produced a real-world dataset, denoted by $\mathbf{S}_{real}$, comprising eight complete RSS maps corresponding to eight different obstacle configurations. The collected measurements capture realistic RF propagation effects, including signal attenuation, multipath reflections, and spatial coverage variations, providing an empirical basis for evaluating the generalization capability and robustness of the \name framework beyond synthetic Digital Twin environments. A representative view of the physical SDR testbed, along with sample RSS maps, is presented in Fig.~\ref{fig:Dataset}(m)--(p).
}

\subsection{DT Construction, Annotation Policy and Detection Targets}
\label{subsec:testbed_digital_twin_creation}
We utilize Blender \cite{blender} exclusively as a geometric modeling engine to construct the spatial DTs of both environments. The physical dimensions, bounding boxes, and fundamental shapes of the walls, ground planes, and scene objects are meticulously modeled. Crucially, each 3D object mesh is mapped to a standardized International Telecommunication Union (ITU) material class (e.g., \texttt{itu\_wood}, \texttt{itu\_glass}, \texttt{itu\_concrete}, \texttt{itu\_metal}). Once the geometric boundaries and material tags are established, the scenes are exported in the \texttt{Mitsuba 3} XML format. This ensures that the structural topology is perfectly preserved when imported into the propagation simulator. Additional details regarding our DT construction methodology can be found in \cite{Sunday2024, Sunday-2024, amatare2025DT}.

\textcolor{blue}{PRISM performs material characterization rather than semantic object recognition: each ground-truth target is a bounding box paired with a material label (e.g., \emph{Wood}, \emph{Metal}), not an object category such as ``chair'' or ``table''. Ground-truth targets comprise the large structural and furniture elements of each scene, i.e., the tables in the Cubicle and Meeting\_Areas, whose footprints span several receiver grid cells. The chairs in both scenarios (12 and 16, respectively) are present in the digital twin, carry ITU material tags, and therefore contribute to the simulated multipath in every propagation map; however, they are \emph{not} annotated as detection targets. This is a deliberate consequence of the spatial resolution of the propagation map: the receiver grid is sampled at approximately 1\,m, and the map is further tokenized into $16{\times}16$ patches covering roughly 3\,m each, whereas a chair footprint is on the order of 0.5\,m. At 3.5\,GHz such objects occupy less than one grid cell and produce a scattering signature that cannot be reliably separated from the surrounding multipath of the adjacent table. Accordingly, Fig.~\ref{fig:Dataset} shows exactly four annotated targets per map. We state this explicitly as a limitation of the current formulation and discuss extensions to small, thin-structured objects in Section~VIII.}

\subsection{Electromagnetic Propagation Modeling}
\label{subsec:testbed_prop_model}

{\color{blue}To model the physical behavior of RF signals, the XML-formatted DTs are imported into NVIDIA's Sionna RT. Sionna parses the ITU material tags assigned during the geometric modeling phase and associates them with frequency-dependent electromagnetic properties. This enables the ray-tracing simulator to model key propagation mechanisms, including specular reflection, transmission through dielectric materials (with refraction), and diffraction around obstacles. In our experiments, specular reflection, refraction, and diffraction are explicitly enabled to capture these propagation effects and account for material-dependent interactions between electromagnetic waves and the surrounding environment.}

To accelerate the intensive ray-tracing computations, we leverage the low-level virtual machine (LLVM) toolchain. The transceiver setup is standardized across all scenes: the Tx is configured as an $8\times 2$ \texttt{tr38901} dual-polarized antenna array \cite{nvidia_rt_details}, and the Rx is an omnidirectional $1\times 1$ dipole planar array. We execute the \texttt{compute\_paths} routine to evaluate the multipath components, bounding the simulation to a maximum interaction depth of 10 and 20 discrete propagation samples per path.

In both environmental scenarios (detailed in Section~\ref{Sec:scenario}), the transmitter is modeled as a drone-mounted mobile node operating near the ceiling \cite{Gerwen2022IndoorDrone, Mao2017IndoorFwDrone}. To empirically emulate the continuous flight mobility of this aerial node, the Tx spatial coordinates are extensively randomized along the upper spatial boundary across discrete simulation instances. To generate the propagation maps, the receive plane is sampled at a height of 1.5 m on a uniform $0.178$ m grid. The primary propagation parameters are summarized in Table~\ref{tab:propagation_params}.



\subsection{End-to-End Synthesis Procedure}
\label{sec:synthesis_algo}
\textcolor{blue}{To make the synthesis pipeline reproducible, Algorithm~\ref{alg:synthesis}
consolidates the steps described in Sections~\ref{Sec:scenario}--\ref{subsec:testbed_prop_model}
into a single procedure. Given a scenario geometry, a material-assignment
distribution, a transmitter sampling distribution, a receiver grid, and a
fixed set of propagation parameters, the procedure returns one annotated
RF propagation map per iteration and assigns it to a dataset split
according to a deterministic rule. All stochastic choices are drawn from
a seeded pseudo-random generator, so every sample in the released dataset
can be regenerated from its recorded seed. The geometry and material tags
of a scene are frozen for a block of $N_{\text{block}}=500$ consecutive
samples (Section~\ref{Sec:scenario}); within a block only the transmitter
position varies. Training-time augmentation (Section~\ref{subsec:experiments-transmitter-mobility})
is applied downstream of this procedure and is not part of the released
maps.}

\begin{algorithm}[!t]
\caption{PRISM RF Propagation Map Synthesis}
\color{blue}
\arrayrulecolor{blue}
\label{alg:synthesis}
\begin{algorithmic}[1]
\REQUIRE Scenario geometry $\mathcal{G}$ (Blender mesh: walls, floor, tables, chairs)
\REQUIRE Material distribution $p(m)$ over $\mathcal{Y}_{\text{known}}=\{\texttt{itu\_wood},\texttt{itu\_glass},\texttt{itu\_concrete},\texttt{itu\_metal}\}$; for the open-set split also $\mathcal{Y}_{\text{unknown}}=\{\texttt{itu\_marble},\texttt{itu\_brick}\}$
\REQUIRE Tx sampling distribution $p(\tau)$ over the ceiling plane at $z_{\text{tx}}=2.8$\,m (uniform for Standard/Cross-Scenario; corner-restricted for Hard Samples)
\REQUIRE Rx coverage-map plane at $z_{\text{rx}}=1.5$\,m: uniform lattice $\mathcal{R}$, cell size $\Delta_{\text{rx}}=0.178$\,m ($7$\,in), yielding $235\times230$ cells
\REQUIRE Propagation parameters $\Theta$: $f_c=3.5$\,GHz, Tx array ($8\times2$ \texttt{tr38901}, dual-pol.), Rx ($1\times1$ dipole), max.\ depth $D=10$, samples per path $S=20$
\REQUIRE Block size $N_{\text{block}}=500$, total $N_{\text{total}}$, split rule $\Pi$ (Table~\ref{tab:split_roles}), seed $s$
\ENSURE $\mathcal{D}=\{(\mathcal{M}_n, Y_n, \text{split}_n)\}_{n=1}^{N_{\text{total}}}$
\STATE Seed PRNG with $s$; $\mathcal{D}\leftarrow\emptyset$
\FOR{$n=1$ \TO $N_{\text{total}}$}
  \IF{$(n-1)\bmod N_{\text{block}}=0$}
    \STATE Sample object positions in $\mathcal{G}$ (non-overlap, wall clearance)
    \STATE For each table $o_j$: sample $m_j\sim p(m)$; open-set split draws three known and one unknown material
    \STATE Assign fixed ITU tags to walls, floor, chairs
    \STATE Export to Mitsuba~3 XML; load into Sionna RT, which maps ITU tags to $(\varepsilon_r,\sigma)$ at $f_c$
  \ENDIF
  \STATE Sample $\tau_n\sim p(\tau)$; place Tx array at $\tau_n$
  \STATE $\mathcal{P}\leftarrow\texttt{compute\_paths}(\mathcal{G},\tau_n,\mathcal{R};D,S)$
  \STATE $G(r)\leftarrow$ path gain [dB] aggregated from $\mathcal{P}$, $\forall r\in\mathcal{R}$ \COMMENT{no additive noise at synthesis; observed range $\approx[-90,-60]$\,dB}
  \STATE Rasterize $G$ on $\mathcal{R}$ ($235\times230$), min--max normalize, resize to $224\times224$ $\Rightarrow\mathcal{M}_n$
  \STATE $Y_n\leftarrow\{(b_j,c_j)\}$: project each table footprint to a normalized box $b_j=(c_x,c_y,w,h)$; $c_j=m_j$ if $m_j\in\mathcal{Y}_{\text{known}}$, else no box (background)
  \STATE $\text{split}_n\leftarrow\Pi(n,\mathcal{G},p(\tau))$; append $(\mathcal{M}_n,Y_n,\text{split}_n)$ to $\mathcal{D}$; log $(s,n,\tau_n,\{m_j\})$
\ENDFOR
\RETURN $\mathcal{D}$
\end{algorithmic}
\end{algorithm}

\textcolor{blue}{\noindent\textbf{Split rule $\Pi$.} Samples generated in the Meeting\_Area
with $p(\tau)$ uniform are assigned to the Standard Split and partitioned
80/10/10 into train/val/test by sample index; those with corner-restricted
$p(\tau)$ form the Hard-Samples split; those containing one
$\mathcal{Y}_{\text{unknown}}$ object form the Open-Set split. All Cubicle-Area
samples form the Cross-Scenario split and are partitioned once, disjointly,
into support (25), validation (75), and test (400) by sample index.}

\subsection{Dataset Taxonomy} 
\label{subsec:dataset-taxonomy}
Following the propagation modeling, the synthesized RF maps are categorized into three distinct subsets to facilitate rigorous model evaluation. A quantitative summary of the dataset is provided in Table~\ref{tab:split_roles}. 


\subsubsection{Mobile Transmitter (Standard Split)} 
\label{subsubsec:mobile-transmitter}

To train and evaluate the detection framework proposed in Section~\ref{subsec:framework-transmitter-mobility}, continuous transmitter mobility is empirically emulated through the randomized placement of the transmitter across the spatial grid. Using this setting, we generate a robust training and testing split comprising $2{,}483$ diverse RF maps exhibiting $-90$ to $-60$ dB signal-to-noise ratios and shadowing effects. A few representative samples are shown in Fig.~\ref{fig:Dataset}(f)-(h) and (j)-(l).


\subsubsection{Hard Samples (Edge Cases Split)} 
\label{subsubsec:hard-samples}
To stress-test localization accuracy under extreme geometric constraints, we synthesize $50$ ``hard samples'' from the in the Meeting\_Area scenario. In these configurations, the transmitter is deliberately placed at the absolute extremities (corners and boundaries) of the scenes. This induces maximal NLoS conditions and severely oblique reflection angles, presenting a significant challenge for spatial feature extraction.  A few samples are shown in Fig.~\ref{fig:Dataset}(k), (l).

\textcolor{blue}{\textit{3) Unseen Scenario (Cross-Scenario Split):} To evaluate the few-shot adaptation described in Section~IV-B, the two scenarios play fixed and non-interchangeable roles throughout the paper. The \textbf{Meeting\_Area} is the \emph{source} domain $\mathcal{S}_{train}$: all 2,483 Standard-Split maps are generated there, and the baseline model never observes a Cubicle-Area map. The \textbf{Cubicle\_Area} is the \emph{target} domain $\mathcal{S}_{new}$: we synthesize 500 maps in this scenario and partition them, once and disjointly, into a 25-sample support set $\mathcal{D}_{support}$ used for fine-tuning, a 75-sample target-validation subset used exclusively to select the class-specific threshold $\delta_{\mathrm{glass}}$, and a 400-sample target-test subset on which all cross-scenario metrics are reported. Representative samples are shown in Fig.~6(b)--(d). Table~\ref{tab:split_roles} summarizes the role of every split.}

\subsubsection{Unknown Materials (Open-Set Split)} 
\label{subsubsec:open-set}
To evaluate the open-set rejection capability described in Section \ref{subsec:framework-open-set}, we synthesize an 
OOD test set with $100$ samples in the Meeting\_Area scenario. In our test set, we consider wood, glass, concrete, and metal {\it known materials},the corresponding material properties in Blender are \texttt{itu\_wood}, \texttt{itu\_glass}, \texttt{itu\_concrete}, \texttt{itu\_metal}, respectively. On the other hand, we consider marble and brick as {\it unknown materials}, corresponding material properties in Blender are \texttt{itu\_marble} and \texttt{itu\_brick}, respectively. While generating each sample of this test set, we randomly include objects with three (out of four) known materials and one (out of two) unknown materials.


\vspace{3pt}
\noindent {\bf Remark 1: } \textit{By establishing a fully deterministic, software-defined data generation pipeline, we successfully overcome the traditional bottleneck of empirical, large-scale RF data collection. The public release of our $3{,}133$-sample dataset, 3D digital twin models, and the complete \name~codebase ensures guaranteed data sufficiency for continuous domain adaptation and reproducible benchmark testing. This explicitly validates Supporting Contribution 2.}

\begin{table}[htbp]
\caption{Sionna RT Electromagnetic Propagation Parameters.}
\begin{center}
\renewcommand{\arraystretch}{1.2}
\begin{tabular}{l l}
\hline
\textbf{Parameter} & \textbf{Configuration Setting} \\
\hline
Simulation Engine & NVIDIA Sionna RT (LLVM accelerated) \\
\textcolor{blue}{Carrier Frequency} & \textcolor{blue}{$3.5$ GHz (Sionna RT default)} \\
Tx Antenna Array & $8 \times 2$ (\texttt{tr38901}), Dual Polarization \\
Rx Antenna & $1 \times 1$ Dipole Planar Array \\
\textcolor{blue}{Tx Height / Rx Plane Height} & \textcolor{blue}{$2.8$ m / $1.5$ m} \\
\textcolor{blue}{Coverage-Map Cell Size} & \textcolor{blue}{$0.178$ m $\times$ $0.178$ m ($7$ in $\times$ $7$ in)} \\
\textcolor{blue}{Rendered Map Size} & \textcolor{blue}{$235 \times 230$ cells (resized to $224 \times 224$)} \\
 \textcolor{blue}{Specular Reflection}&\textcolor{blue}{True}\\
 \textcolor{blue}{Diffraction}&\textcolor{blue}{True}\\
 \textcolor{blue}{Refraction}&\textcolor{blue}{True}\\
Ray Tracing Routine & \texttt{compute\_paths} \\
Maximum Interaction Depth & $10$ \\
Propagation Samples & $20$ \\
\textcolor{blue}{Scene Reconfiguration Period} & \textcolor{blue}{Every $500$ maps} \\
\textcolor{blue}{Additive Noise at Synthesis} & \textcolor{blue}{None (augmentation only, Section~VI-B1)} \\
\hline
\end{tabular}
\label{tab:propagation_params}
\end{center}
\end{table}


\begin{table*}
\color{blue}
\centering
\caption{\color{blue}Roles of the PRISM dataset splits in each experiment. All Cubicle-Area samples are held out from source training; the 25-sample support set, the validation subset used for threshold selection, and the test subset are mutually disjoint.}
\label{tab:split_roles}
\small
\begin{tabular}{llrl}
\toprule
\textbf{Split} & \textbf{Scenario} & \textbf{\#Samples} & \textbf{Role} \\
\midrule
\multicolumn{4}{l}{\emph{Source domain} $\mathcal{D}_{source}$ (Exp.~1, Section~VI-B)} \\
\quad Standard -- train & Meeting\_Area & 1,987 & Baseline training \\
\quad Standard -- val & Meeting\_Area & 248 & Early stopping / monitoring \\
\quad Standard -- test & Meeting\_Area & 248 & In-domain evaluation (Section~VII-A) \\
\midrule
\multicolumn{4}{l}{\emph{Target domain} $\mathcal{D}_{target}$ (Exp.~2, Section~VI-C)} \\
\quad Cross-Scenario -- support $\mathcal{D}_{support}$ & Cubicle\_Area & 25 & Few-shot fine-tuning of DETR head \\
\quad Cross-Scenario -- val & Cubicle\_Area & 75 & Selection of $\delta_{\mathrm{glass}}$ only \\
\quad Cross-Scenario -- test & Cubicle\_Area & 400 & Reported target metrics (Section~VII-B, Fig.~9) \\
\midrule
\multicolumn{4}{l}{\emph{Stress tests on the source layout}} \\
\quad Hard Samples (Edge Cases) & Meeting\_Area & 50 & Extreme Tx placement (Section~VII-B3) \\
\quad Open-Set & Meeting\_Area & 100 & Unknown-material rejection (Section~VII-C) \\
\bottomrule
\end{tabular}
\end{table*}

%% file: Sections/experiments-new.tex
\section{Experiments}
\label{sec:experiments}

In this section, we orchestrate various experimental settings to validate the proposed framework presented in Section \ref{sec:proposed_framework}.
\subsection{Implementation Details and Hyperparameters}
\label{subsec:exp_setup}
We implement proposed the \name~framework using PyTorch and train all models on a dedicated NVIDIA A100 GPU computing cluster.  

The ViT backbone is configured with $L = 12$ Transformer encoder layers. Input RF propagation maps are resized to spatial dimensions of $H = 224$ and $W = 224$, and partitioned into patches of size $P = 16$. This yields a flattened sequence of $N_{tokens} = 197$ tokens (inclusive of the \texttt{[CLS]} token), which are projected into an embedding space of $d_{model} = 768$. For the DETR-style set prediction head, we empirically set the number of object queries to $N_q = 50$, which serves as an upper bound for the maximum number of objects detectable in a single scene. The model is trained to classify $K = 4$ known material categories (\textit{Concrete, Glass, Metal, Wood}), plus an explicitly modeled \texttt{background} class. 

\noindent{\bf Evaluation Metrics:} To evaluate detection performance, we adopt standard ML localization metrics: Precision ($\mathcal{P}$), Recall ($\mathcal{R}$), and the F1 Score ($\mathcal{F}_1$). Precision quantifies the proportion of predicted positive detections that are strictly correct, while recall measures the proportion of ground-truth physical targets successfully identified. The F1 score ($2 \times \frac{\mathcal{P} \times \mathcal{R}}{\mathcal{P} + \mathcal{R}}$) provides a balanced assessment of characterization quality, penalizing both false positives and false negatives.

{\color{blue}\noindent \textbf{Adaptation cost.} Table~\ref{tab:adapt_cost} contrasts the two
training stages. Because the ViT backbone $\psi$ is frozen, the 25-shot
stage updates only the DETR head $\phi$ and converges in 50 iterations,
requiring under 42\,s of wall-clock time and under 3.8\,GB of peak VRAM on
the same A100 used for baseline training. This is roughly three orders of
magnitude cheaper than the 4.5\,h baseline stage. We emphasize that these
figures characterize adaptation at the MEC server, where the digital twin
and propagation maps already reside.}

\subsection{Experiment 1: Transmitter Mobility-Agnostic Detection}
\label{subsec:experiments-transmitter-mobility}
\subsubsection{Dataset and Augmentation Pipeline}
To evaluate the spatial robustness of the ViT-DETR architecture under varying transmitter locations, we utilize the \textit{Standard Split} comprising $2{,}483$ RF maps. To rigorously prevent data leakage, we partition this dataset into training, validation, and testing sets using an $80/10/10$ split ratio. 

To prevent overfitting on the limited sample size and force the model to learn geometry-invariant electromagnetic features, we implement a comprehensive stochastic augmentation pipeline. Each original training sample undergoes transformations generating three augmented instances. The pipeline includes random horizontal and vertical flips, random crop-and-zoom operations (scaling between $10\%$ and $30\%$), random affine rotations ($\pm 15^\circ$), and signal intensity perturbation via saturation adjustments ($\pm 25\%$).

\begin{figure*}[t]
    \centering
    \includegraphics[width=\textwidth]{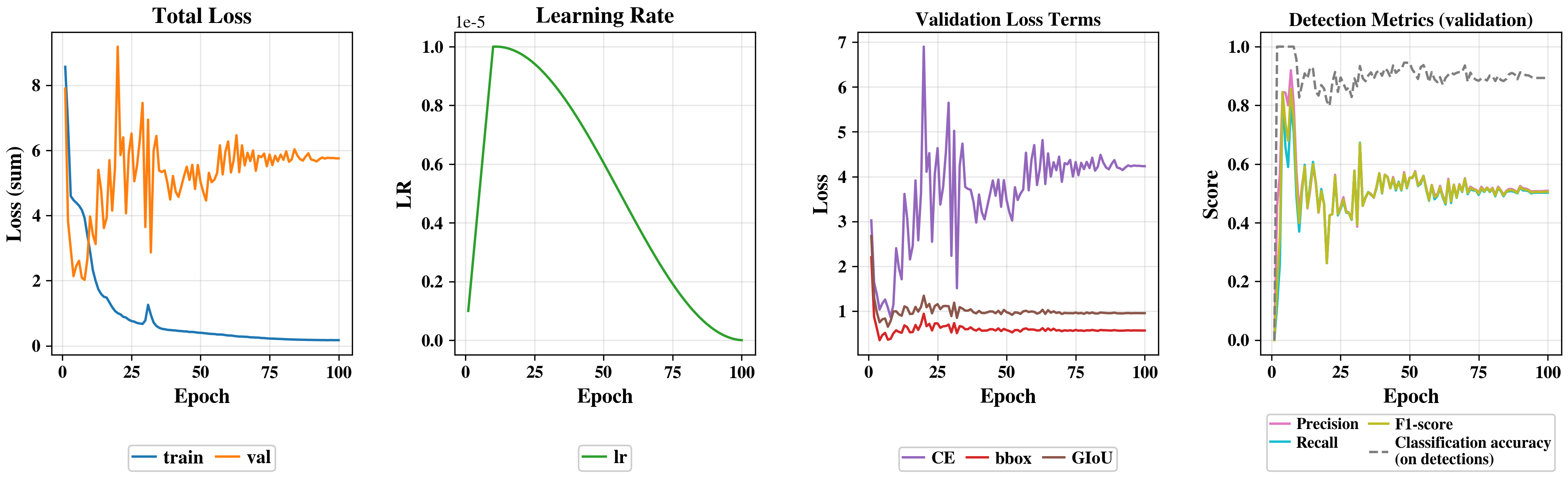}
    \caption{Training dynamics of the proposed \name~architecture over 100 epochs. 
    (Leftmost) Total weighted bipartite matching loss for training and validation splits. 
    (Middle-left) The hybrid learning rate schedule featuring a linear warmup to $1\times10^{-5}$ followed by cosine annealing. 
    (Middle-right) Decomposed validation loss components: cross-entropy classification ($\mathcal{L}_{ce}$), bounding box regression ($\mathcal{L}_{L1}$), and GIoU loss ($\mathcal{L}_{giou}$). 
    (Rightmost) Detection metrics on the validation set, tracking precision, recall, F1 score, and classification accuracy for matched queries.}
    \label{fig:training}
\end{figure*}

\subsubsection{Training Dynamics and Convergence}
The training behavior of \name~over $100$ epochs is documented in Fig.~\ref{fig:training}. The network weights are optimized using a hybrid learning rate schedule: a linear warmup phase scales the learning rate to a peak of $1\times 10^{-5}$ at epoch $10$, subsequently followed by monotonic cosine annealing to zero. 

As depicted in the decomposed loss analysis, the spatial bounding box regression losses ($\mathcal{L}_{L1}$ and $\mathcal{L}_{giou}$) converge rapidly and exhibit high stability, indicating that the Transformer cross-attention effectively extracts localized spatial coordinates. Conversely, the cross-entropy classification loss ($\mathcal{L}_{ce}$) accounts for the majority of the total validation loss manifold, stabilizing at approximately $4.2$. This dynamic implies that distinguishing subtle material scattering signatures—rather than spatial localization—constitutes the primary optimization bottleneck. The observable train-validation gap is a direct artifact of the highly constrained validation split size ($10\%$). Consequently, we rely on the out-of-distribution evaluation benchmark (Section \ref{subsec:exp2_few_shot}) as the definitive measure of generalized performance.

\subsection{Experiment 2: Few-Shot Cross-Scenario Generalization}
\label{subsec:exp2_few_shot}
\textcolor{blue}{To validate cross-scenario adaptability, we use the \textit{Cross-Scenario Split} defined in Section~V-E3: the baseline model, trained only on Meeting-Area maps, is deployed in the \textbf{Cubicle\_Area}, whose enclosed corridors and dense NLoS geometry differ substantially from the open-plan source layout (Section~V-A). Performance is quantified using $\mathcal{P}$, $\mathcal{R}$, and $\mathcal{F}_1$ on the 400-sample target-test subset after fine-tuning the DETR head on the 25-sample support set. The Hard-Samples evaluation below is a separate stress test on the \emph{source} layout and does not involve adaptation.}

Furthermore, to stress-test spatial invariant feature extraction, we evaluate the architecture against the curated $50$ \textit{Edge Cases Split}.  These extreme-edge transmitter placements introduce severe line-of-sight occlusions, near-grazing signal incidence angles, and localized multi-path signal collapse. Despite these deliberate degradations, \name~maintains strict detection fidelity, empirically validating that the learned ViT embeddings abstract the fundamental electromagnetic properties of the materials rather than memorizing environment-specific transmission paths.

\subsection{Experiment 3: Open-Set Unknown Material Rejection}
\label{subsec:experiments-open-set}
In safety-critical MEC deployments, anomalous materials must be actively suppressed rather than erroneously classified. To evaluate open-set rejection, we use the \textit{Open-Set Split} with $100$ OOD samples containing \textit{Marble} and \textit{Brick} anomalies.
In these tests, the model must simultaneously localize the three known objects while successfully ignoring the spatial footprint of the unknown material. 

\begin{figure}
    \centering
    \includegraphics[width=1\linewidth]{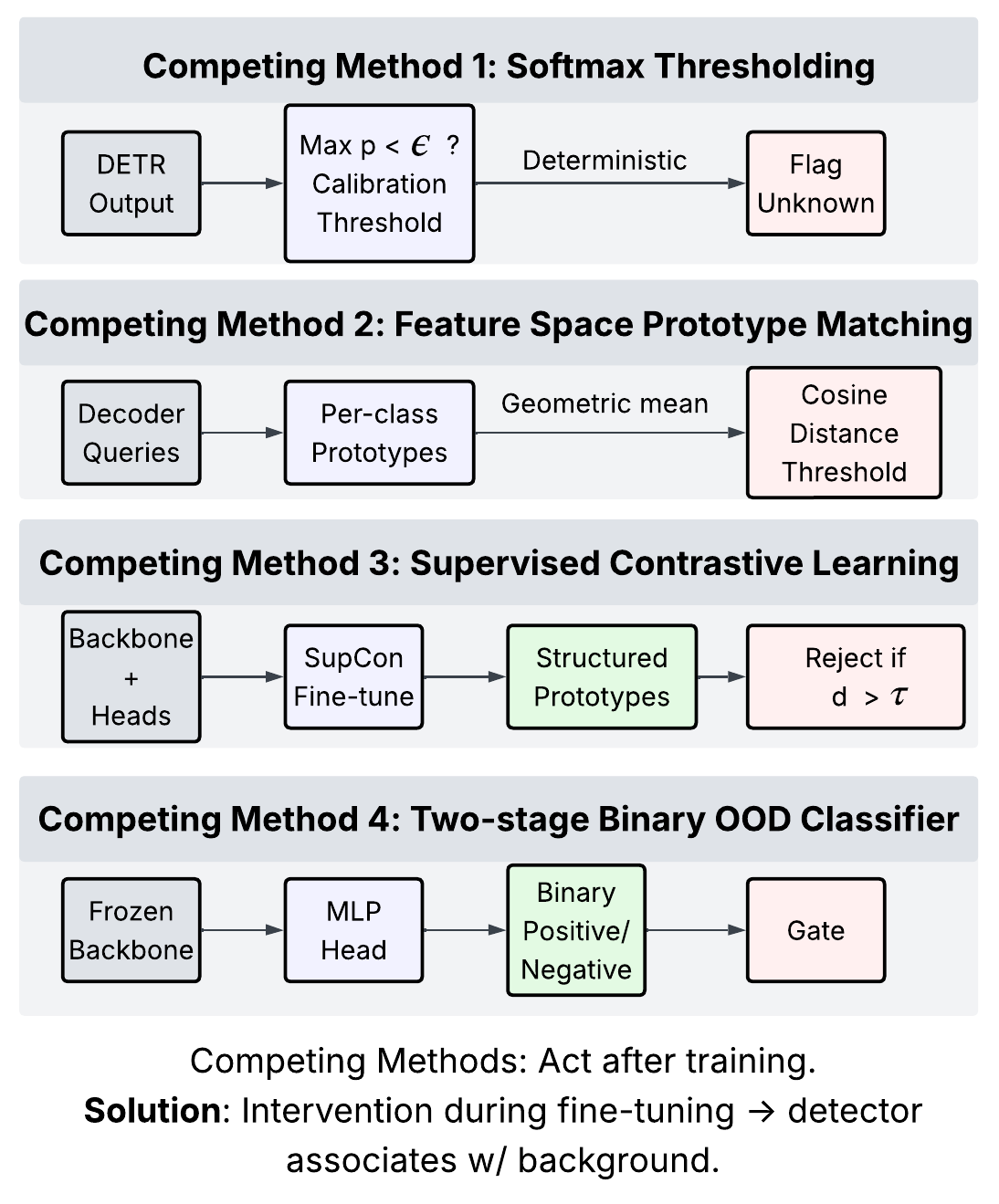}
    \caption{Various competing methods for open-set unknown material rejection experiments.}
    \label{fig:competing}
\end{figure}

We systematically evaluate our proposed outlier exposure method with four competing rejection strategies (shown in Fig.~\ref{fig:competing}), measuring the unknown suppression rate against the collateral false-rejection rate of known materials:

\subsubsection{Competing Method 1: Softmax Score Thresholding}
A post-hoc deterministic baseline. A detection is flagged as \textit{Unknown} if the maximum probability across the softmax distribution for all known classes falls below a calibrated threshold $\delta$.

\subsubsection{Competing Method 2: Feature-Space Prototype Matching}
Operating within the DETR decoder's internal representation space, we aggregate Hungarian-matched query embeddings to construct per-class geometric mean prototypes. During inference, cosine similarity distances between the output query embeddings and these prototypes are computed; queries exceeding a maximum distance threshold are suppressed.

\subsubsection{Competing Method 3: Supervised Contrastive Learning (SupCon)}
We introduce an auxiliary $128$-dimensional projection head to the backbone. The model is jointly fine-tuned using a Supervised Contrastive Loss ({\tt SupConLoss}) to strictly enforce intra-class feature compactness and inter-class spatial separation. Prototypes are then recomputed in this highly structured projection space for rigorous threshold-based rejection.

\subsubsection{Competing Method 4: Two-Stage Binary OOD Classifier}
A structural modification where a frozen detection backbone feeds into a lightweight binary MLP head ($768 \rightarrow 256 \rightarrow 128 \rightarrow 1$). Decoder features matched to ground-truth bounding boxes are labeled as \textit{positive} (known), while unmatched queries are labeled \textit{negative} (anomalous). This binary classifier acts as an independent post-hoc gating mechanism.

\subsubsection{Proposed Method: Outlier Exposure}
Unlike the post-hoc baseline configurations, this strategy actively intervenes during the $k$-shot fine-tuning phase. The \textit{Brick} and \textit{Marble} anomalies are explicitly injected into the support set, but their bounding boxes are hard-coded to the DETR \texttt{background} class. This theoretically superior approach forces the cross-attention layers to natively associate anomalous electromagnetic scattering variances with the absence of a target, achieving robust anomaly rejection without requiring any external architectural bloat.

%% file: Sections/evaluation-new.tex
\section{Evaluation}
\label{sec:evaluation}

In this section, we present a comprehensive performance evaluation of the \name~framework on the experiments mentioned in Section~\ref{sec:experiments}.


\subsection{Evaluation of Transmitter Mobility-Agnostic Detection}
\label{sec:valid}

\subsubsection{Validation Dynamics}
During the training phase detailed in Section \ref{subsec:exp_setup}, the classification accuracy on matched detections stabilizes between $0.88$ and $0.90$. However, the validation F1 score converges to approximately $0.50$. This artificially suppressed validation score is attributable to two methodological factors: the highly constrained validation split size ($10\%$ of the training corpus, leading to high metric variance) and the absence of post-processing calibration. Specifically, no non-maximum suppression (NMS) or class-specific confidence thresholds are applied during the raw validation monitoring, leading to expected geometric redundancies. 

\subsubsection{Test Set Generalization}
We evaluate the definitive baseline detection performance on the held-out test set drawn from the {\it Standard Split} (Section \ref{subsubsec:mobile-transmitter}). After applying standard inference calibration, the proposed model achieves an F1 score of $0.9201$, a precision of $0.9495$, and a recall of $0.8925$. These robust metrics demonstrate that the ViT-DETR architecture effectively extracts spatially invariant electromagnetic scattering signatures. 

\begin{figure}
    \centering
    \includegraphics[width=1\linewidth]{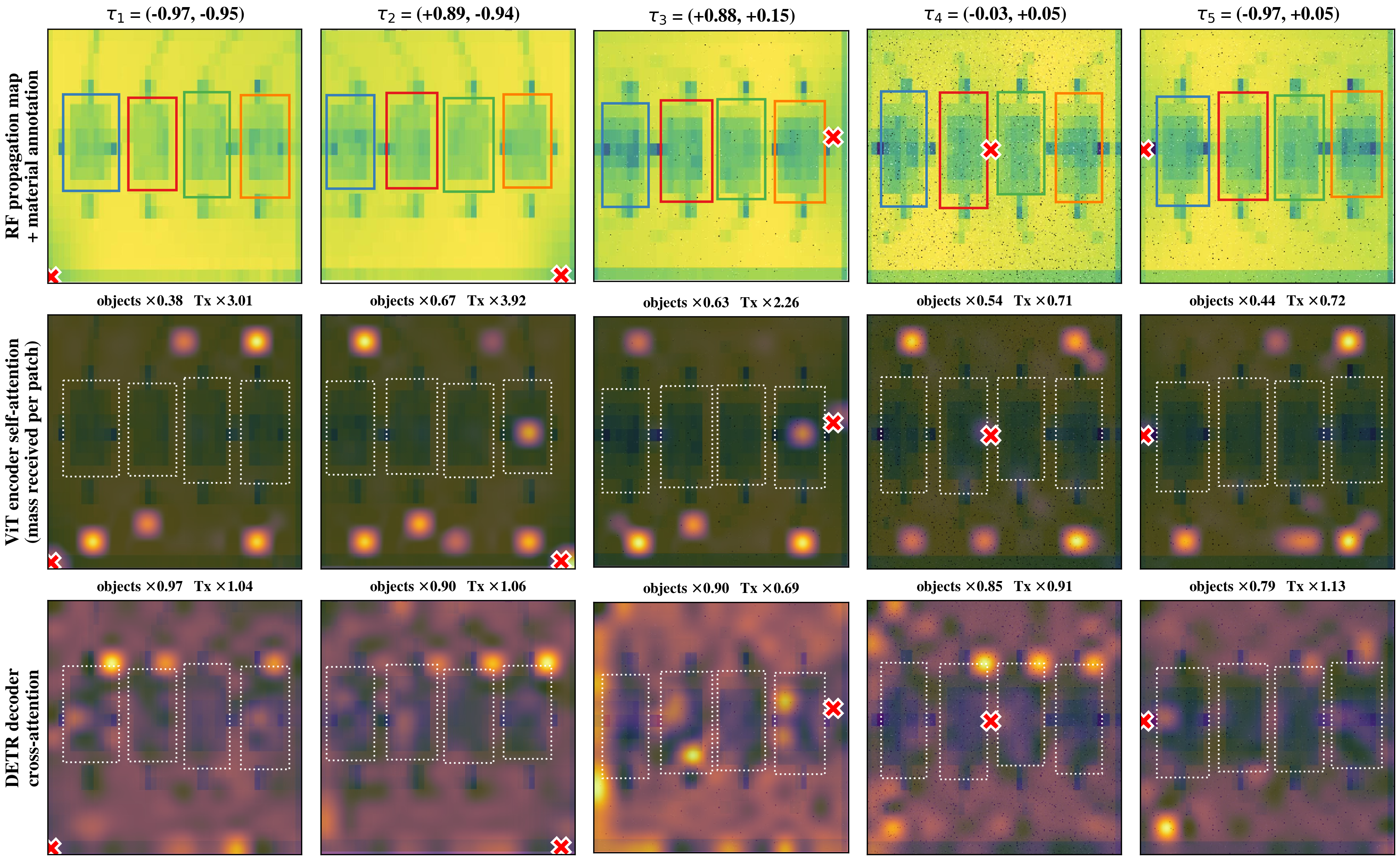}
    \caption{\color{blue}Attention for a single Meeting-Area scene under five transmitter
positions $\tau_1$--$\tau_5$ (red crosses; coordinates normalized to
$[-1,1]$). Top: RF propagation map with material annotations. Middle:
ViT encoder self-attention, shown as the total attention mass received
by each patch, averaged over heads and layers. Bottom: DETR decoder
cross-attention from the matched object queries. Numbers above each panel
give the attention mass on the annotated objects and on the transmitter
patch relative to a uniform baseline ($1.0\times$).}
    \label{fig:attn_tau}
\end{figure}

{\color{blue} \subsubsection{Transmitter Invariance}
\label{sec:attn}
Fig.~\ref{fig:attn_tau} visualizes attention for one Meeting-Area scene
under five transmitter positions spanning the corners, edges, and centre
of the floor. For each map we report the attention mass received by the
annotated object regions and by the patch containing the transmitter,
normalized by the uniform-attention baseline. Two patterns are consistent
across all five $\tau$.

First, the ViT encoder does \emph{not} abstract the transmitter away.
When the transmitter lies at the boundary ($\tau_1$--$\tau_3$), its patch
receives 2.3--3.9$\times$ the uniform attention mass, while the object
regions receive only 0.4--0.7$\times$. The remaining encoder attention
concentrates on the chairs flanking each table, i.e., the small,
high-contrast scatterers that shape the multipath but are not detection
targets (Section~\ref{subsec:testbed_digital_twin_creation}). The encoder therefore encodes the
illumination geometry explicitly, exactly the information a
material-dependent scattering signature must be interpreted against.

Second, the DETR decoder's cross-attention is where invariance appears.
Attention from the object queries to the backbone features is nearly
uniform over the annotated objects (0.79--0.97$\times$) and does not
track the transmitter (0.69--1.13$\times$, i.e., statistically
indistinguishable from uniform) regardless of $\tau$. The boxes and
material labels are thus read out from the same spatial regions whether
the transmitter is at a corner or between the tables.


\vspace{3pt}
\noindent \textbf{Observation 1:} \textit{Trained on randomized and tested on
held-out transmitter positions, PRISM localizes and characterizes
materials regardless of signal origin. Attention analysis
(Section~\ref{sec:attn}) shows this invariance arises in the decoder
readout, while the encoder explicitly represents the transmitter position
as context. This validates Primary Contribution~1.}}

\subsection{Evaluation of Cross-Scenario Generalization}
\label{sec:eval_bench_results}

\subsubsection{Few-Shot Adaptation on Unseen Benchmarks}
To evaluate the domain adaptation strategy introduced in Section \ref{subsec:framework-few-shot}, we deploy the model on the Cross-Scenario target-test subset (400 samples). This environment presents architectural configurations completely unseen during baseline training. Utilizing a minimal $25$-sample support set $\mathcal{D}_{support}$ for lightweight fine-tuning, the model rapidly adapts to the new domain's multipath distribution without overfitting. 

\begin{table}[!t]
\color{blue}
\centering
\caption{\color{blue}Training cost of the baseline stage versus the 25-shot adaptation stage. Both measured on a single NVIDIA A100 (40\,GB); the adaptation numbers therefore characterize MEC-server-side cost, not on-device edge cost.}
\label{tab:adapt_cost}
\footnotesize
\setlength{\tabcolsep}{4pt}
\begin{tabular}{lcc}
\toprule
 & \textbf{Baseline training} & \textbf{25-shot adaptation} \\
\midrule
Data                & 1{,}987 maps ($\times$4 w/ aug.) & 25 maps (no aug.) \\
Trainable modules   & ViT + DETR head              & DETR head only (ViT frozen) \\
Trainable params    & 135M                         & 5M \\
Optimization        & 100 epochs                   & 50 iterations \\
Learning rate       & warmup $\to 10^{-5}$, cosine & 1$\times10^{-5}$, const. \\
Wall-clock time     & $\approx$4.5\,h              & $<$42\,s \\
Hardware            & A100                         & A100 \\
\bottomrule
\end{tabular}
\end{table}

\textcolor{blue}{\subsubsection{Effect of Validation-Set Calibration.}
Before class-specific confidence calibration, the adapted model achieves an overall F1 score of 0.8802 on the target test set. The dominant error source is Glass, for which overlapping transmission and reflection patterns generate a comparatively high number of false positives. We therefore select $\delta_{\mathrm{glass}}=0.45$ exclusively on the independent target-validation subset; no target-test label is used for threshold selection. Applying this fixed threshold to the untouched target-test subset increases Glass F1 from 0.6964 to $0.9937$ and the overall F1 from $0.8802$ to $0.9822$. We report both values to distinguish the intrinsic output of the adapted detector from the performance obtained after validation-set calibration.}

\begin{figure}[t!]
    \centering
    \includegraphics[width=1\linewidth]{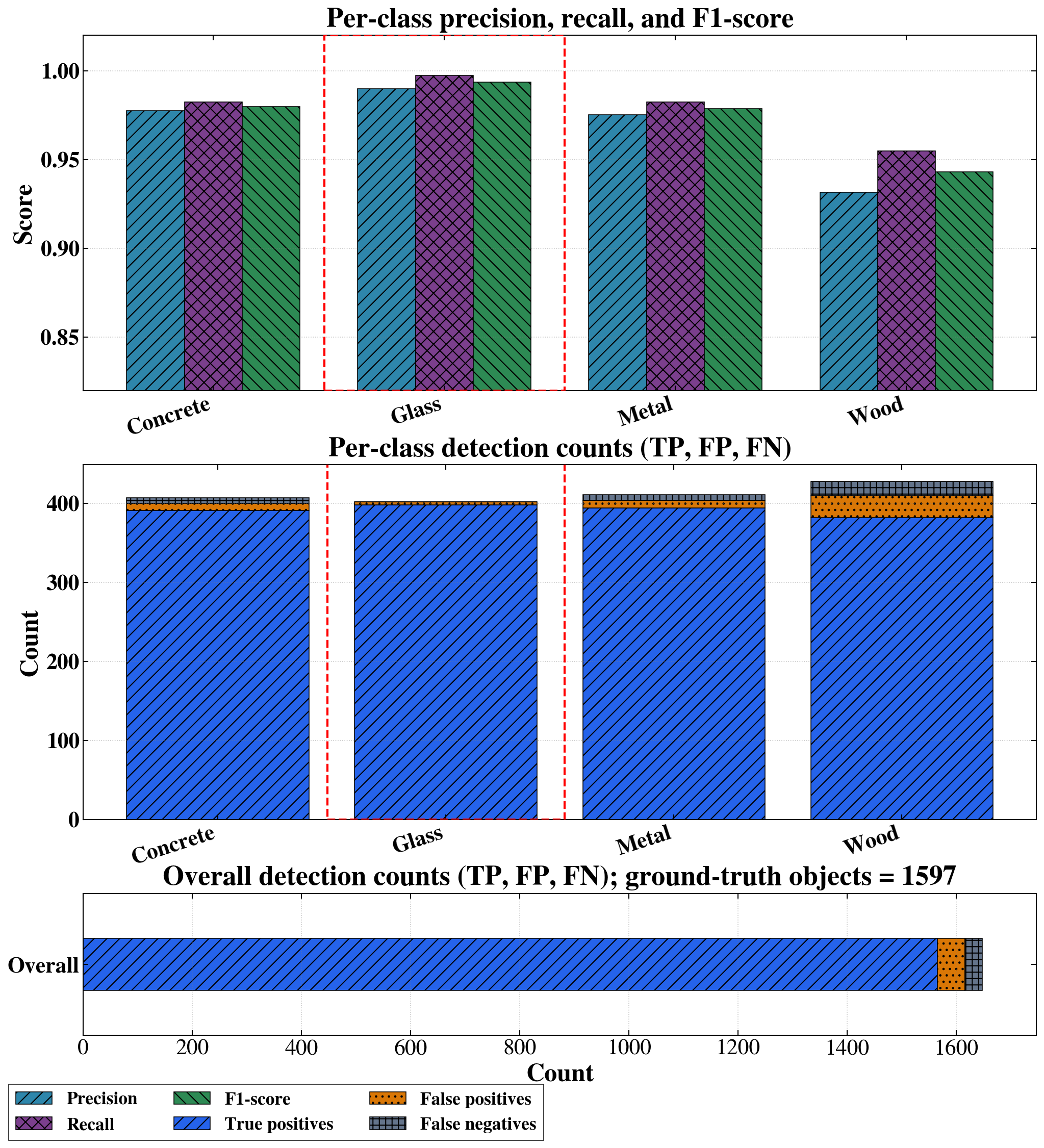}
    \caption{Per-class and overall detection performance on the out-of-distribution evaluation benchmark following few-shot adaptation and threshold calibration(Cubicle\_Area, 400 target-test samples).}
    \label{fig:final}
\end{figure}

\vspace{3pt}
\noindent {\bf Observation 2: } \textit{The successful domain transfer and exceptional F1 score recovery via a minimal support set $\mathcal{D}_{support}$ demonstrates extreme adaptability to novel architectural layouts. This explicitly validates Primary Contribution 2.}

\subsubsection{Robustness under Adverse Line-of-Sight (LoS) Conditions}
\label{subsec:challenging}
\begin{figure}[t!]
	\centering
	\begin{subfigure}{0.45\textwidth}
        \centering        
        \includegraphics[width=\linewidth]{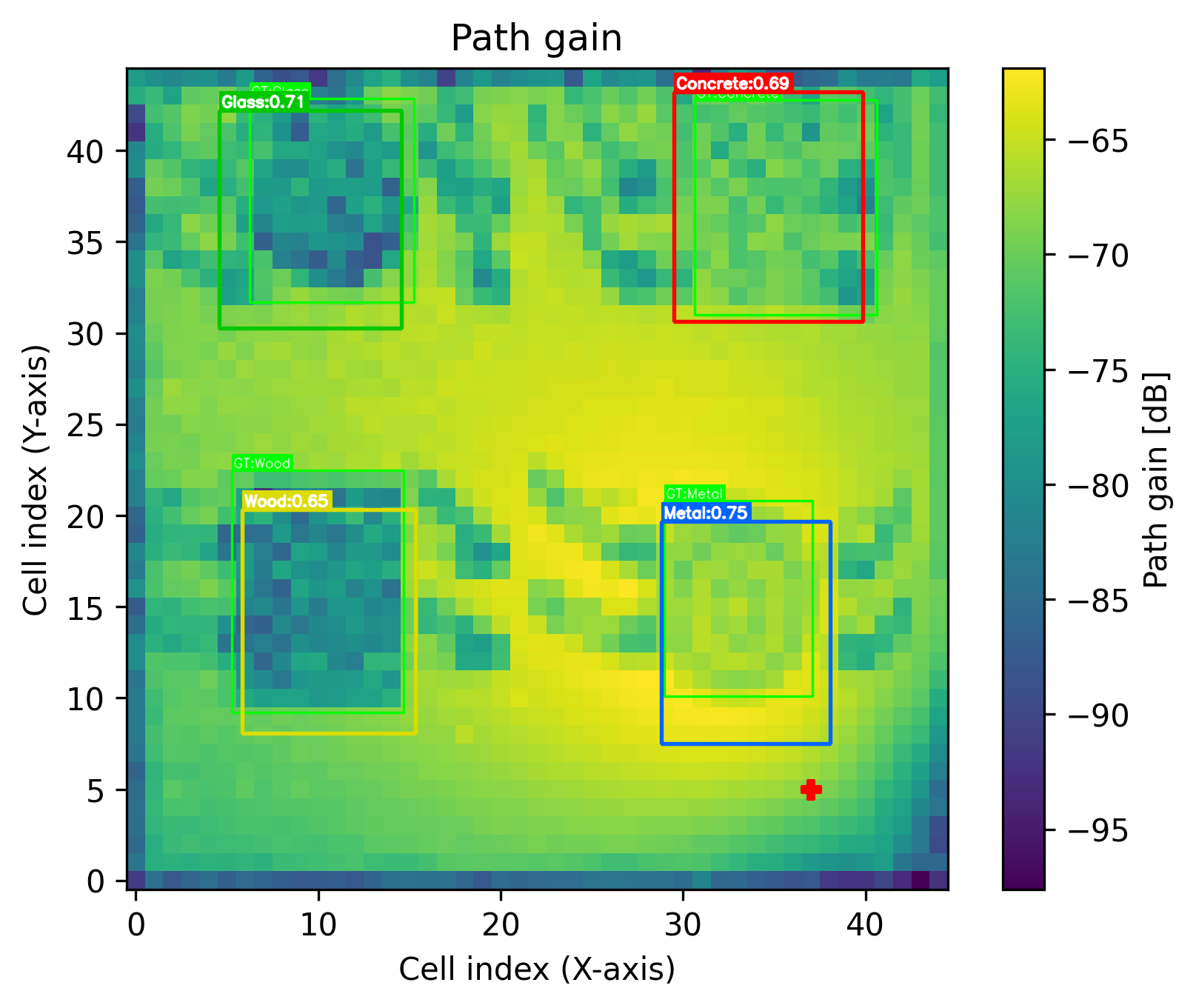}
		\caption{}
	\end{subfigure}
	\begin{subfigure}{0.45\textwidth}
        \centering
        \includegraphics[width=\linewidth]{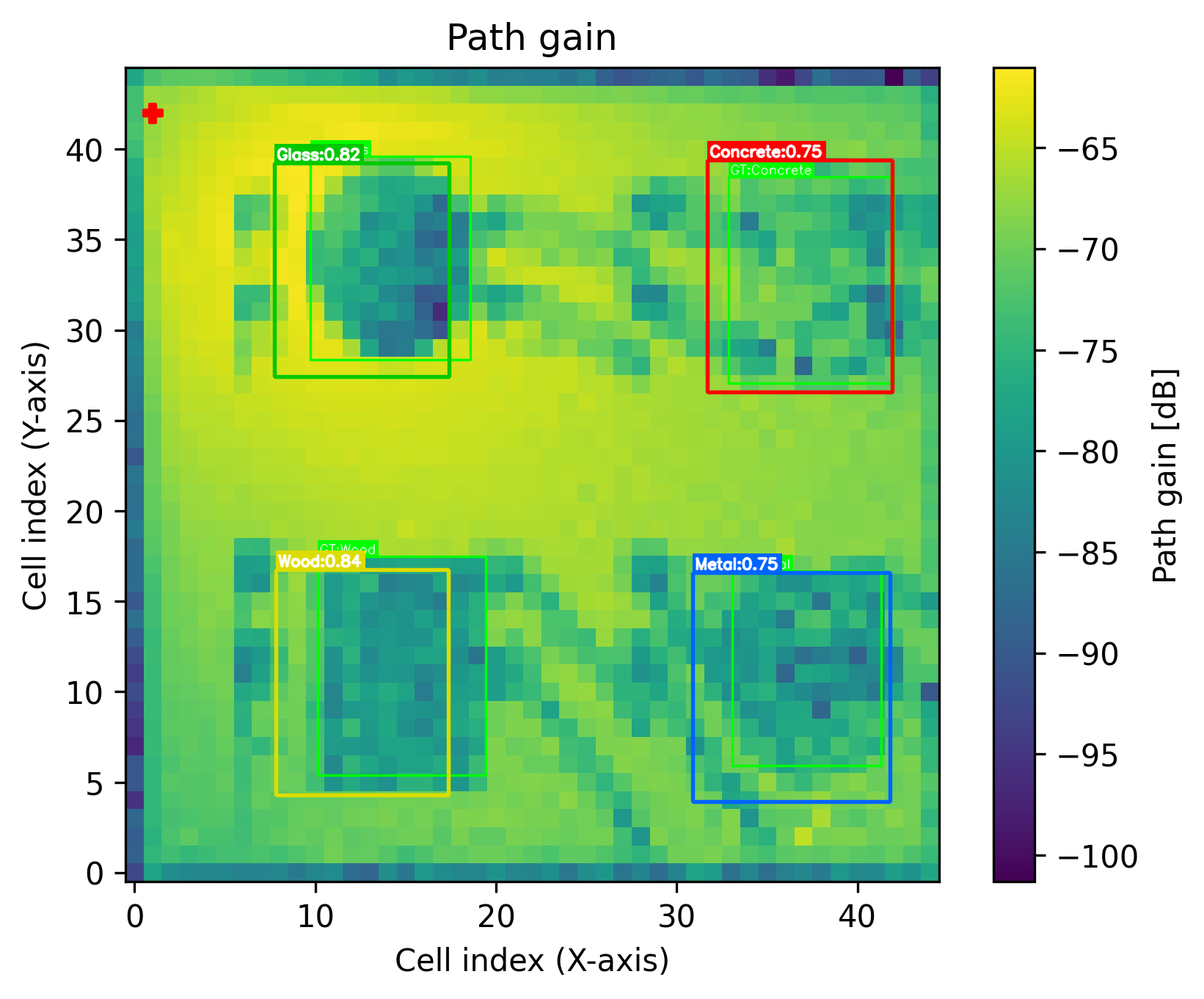}
        \caption{}
	\end{subfigure} 
  \caption{Representative detection outputs on \textit{Hard Samples} exhibiting severely degraded LoS conditions and extreme-edge transmitter placements. The mobile transmitter's positions are at the corners, marked in red cross sign.}
    \vspace{-0.2in}
	\label{fig:hard-samples}
\end{figure}

We stress-test the adapted model against the $50$ samples of the \textit{Edge Cases Split}, presented in Section \ref{subsubsec:hard-samples}. These samples feature near-grazing incidence angles, severe multipath interference, and highly degraded LoS geometries. Despite these deliberate signal obfuscations, \name~maintains strong spatial localization, achieving a mean intersection-over-union (IoU) of $0.86$. Note that while the GIoU is utilized as a differentiable training objective (as defined in Eq.~\ref{eq:c-match} and~\ref{eq:l-total}) to penalize non-overlapping predictions and provide directional gradients, we report the standard IoU here as the universally accepted inference metric for evaluating final spatial overlap accuracy. Representative detection outputs, visualized in Fig.~\ref{fig:hard-samples}, confirm that the learned ViT embeddings are resilient to violated LoS assumptions.

\vspace{3pt}
\noindent {\bf Observation 3: } \textit{The model's sustained high spatial localization accuracy (mean IoU of $0.86$) despite near-grazing incidence angles and severe multipath interference explicitly validates Supporting Contribution 3.}

\subsection{Evaluation of Open-Set Unknown Material Rejection}

To address open-world deployment constraints, we evaluate the system against the {\it Open Set Split} test sets. The model must preserve the detection of the three known objects in the scene while actively suppressing the unknown material anomaly. A comparative summary of the five rejection strategies (formalized in Section \ref{subsec:experiments-open-set}) is provided in Table~\ref{tab:unknown_material_summary}. 

\begin{table*}[t!]
\color{blue}
\centering
\caption{\color{blue}Comparative Analysis of Unknown Material Rejection Strategies.}
\label{tab:unknown_material_summary}
\renewcommand{\arraystretch}{1.2}
\begin{tabular}{l c c l}
\toprule
\textbf{Rejection Approach} & \textbf{Unknown Suppressed} & \textbf{Known Preserved} & \textbf{Bounding Box Quality} \\
\midrule
Softmax Thresholding & 21\% & 64\% & Maintained \\
Feature-Space Prototype & N/A & 33\% (Severely Degraded) & Maintained \\
Supervised Contrastive Prototypes & 50\% & 28\% (Severely Degraded) & Destroyed (IoU $\approx$ 0) \\
Two-Stage Binary OOD Head & 49\% & 38\% & Degraded \\
Outlier Exposure (Box Matching) & 43\% (At correct location) & 58\% & Maintained (Known) \\
\textbf{Proposed: Outlier Exposure (Suppression Mode)} & \textbf{100\%} & \textbf{100\%} & \textbf{Maintained} \\
\bottomrule
\end{tabular}
\end{table*}

\subsubsection{Limitations of Post-Hoc Baselines}
The evaluation reveals that although the traditional post-hoc OOD methods are highly successful in optical computer vision, drastically fail in the RF domain. Unlike RGB imagery where \textit{Brick} and \textit{Concrete} are visually distinct, their RF path gain signatures exhibit substantial multipath overlap. Consequently, Softmax Thresholding forces an unacceptable trade-off, incorrectly rejecting $36\%$ of valid known materials just to catch $21\%$ of the anomalies. Similarly, Supervised Contrastive Learning destroys the bounding box regression head (IoU drops to near zero), proving that joint contrastive optimization structurally conflicts with the DETR spatial localization objective.

\subsubsection{Success of Outlier Exposure}
\begin{figure}[t!]
    \centering
    \includegraphics[width=1\linewidth]{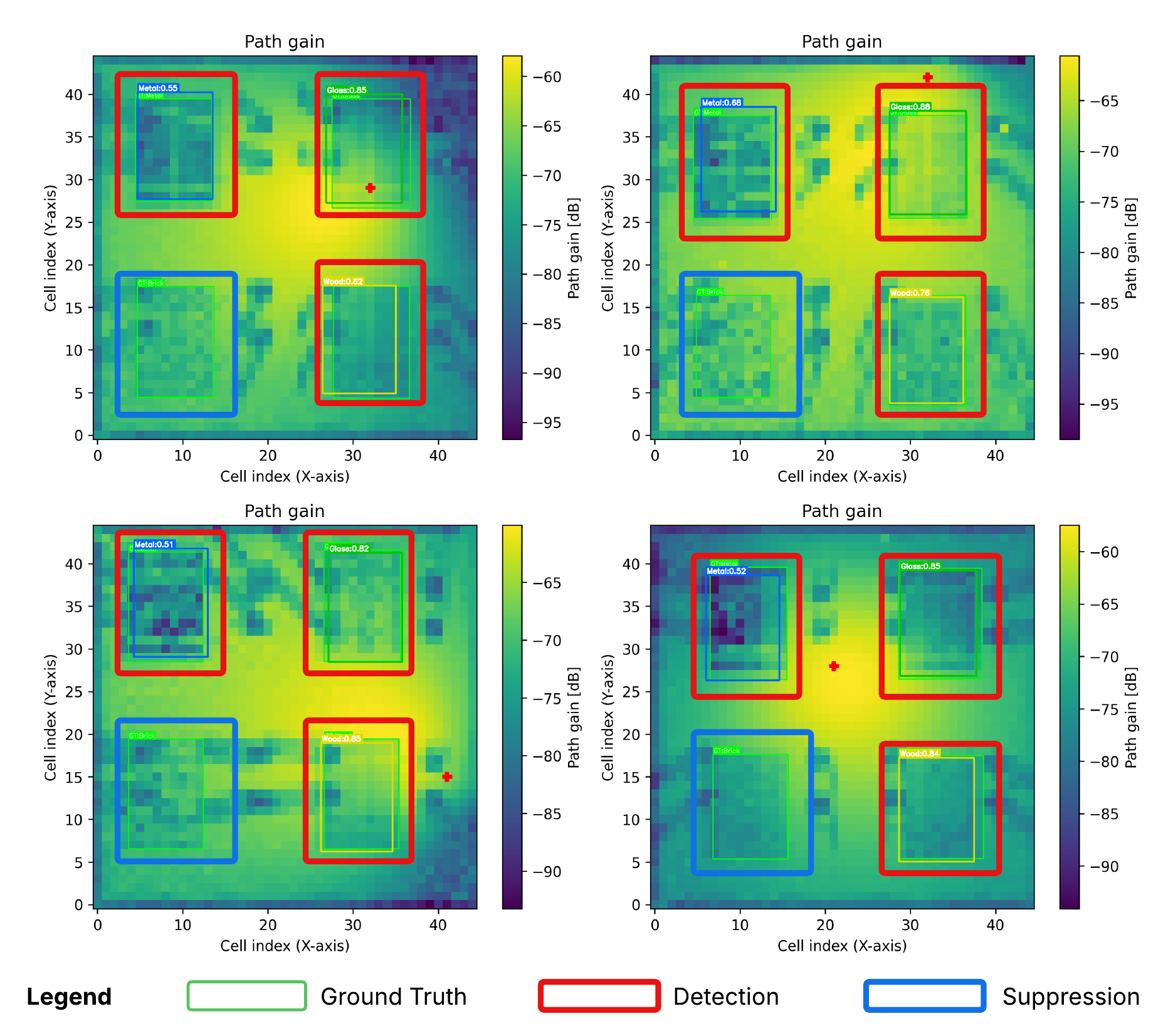}
    \caption{Inference results operating under the proposed Outlier Exposure strategy. The network accurately localizes known materials while completely suppressing bounding box generation for the out-of-distribution \textit{Brick} anomalies, eliminating false positives.}
    \label{fig:Unknown}
\end{figure}

The proposed Outlier Exposure strategy (operating in suppression mode with $\delta = 0.45$) vastly outperforms all alternative architectures. By explicitly injecting OOD samples as \texttt{background} targets during fine-tuning, the model learns to associate anomalous scattering with an absence of a target. Under this setting, the model achieves a $100\%$ suppression rate for the unknown \textit{Brick} objects while correctly detecting an average of $2.8$ known objects per image. Known-class confidence remains strictly robust (Glass: $\mu = 0.527$, Metal: $\mu = 0.535$, Wood: $\mu = 0.487$). As visualized in Fig.~\ref{fig:Unknown}, the model avoids generating false positives for the unseen materials without compromising general localization fidelity.


\vspace{3pt}
\noindent {\bf Observation 4: } \textit{By utilizing outlier exposure to force anomalous signatures into the background class, \name~prevents dangerous false-positive classifications in dynamic environments. This validates Primary Contribution 3.}

{\color{blue}
\subsection{Sim-to-Real Co-Training}
\label{sec:co_training}

To evaluate the sim-to-real generalization of PRISM,
we collected ten real-world propagation maps.
However, acquiring such measurements is time-consuming
and costly, as each map requires a complete robotic sweep
under a fixed transmitter placement (Section~V-B).
We therefore investigate whether synthetic maps from
the digital twin can complement scarce measurements
to improve site-specific detection.

\noindent\textbf{Experimental Setup.}
We train a lightweight PRISM with a ViT-Tiny encoder (5.7M
parameters), retaining the detection head and loss in
Section~IV. Three training sets are compared:
(i) 8 measured maps, (ii) 40 synthetic maps, and
(iii) their combination. All models are trained from scratch
for 3,000 steps with flip augmentation.

Evaluation uses 2 held-out measured maps (8 objects),
collected after the training protocol was fixed, and
146 unseen synthetic maps (584 objects). A detection is
correct if its bounding box achieves IoU $\geq 0.5$
and its predicted material matches the ground truth.

\begin{figure}
    \centering
    \includegraphics[width=1\linewidth]{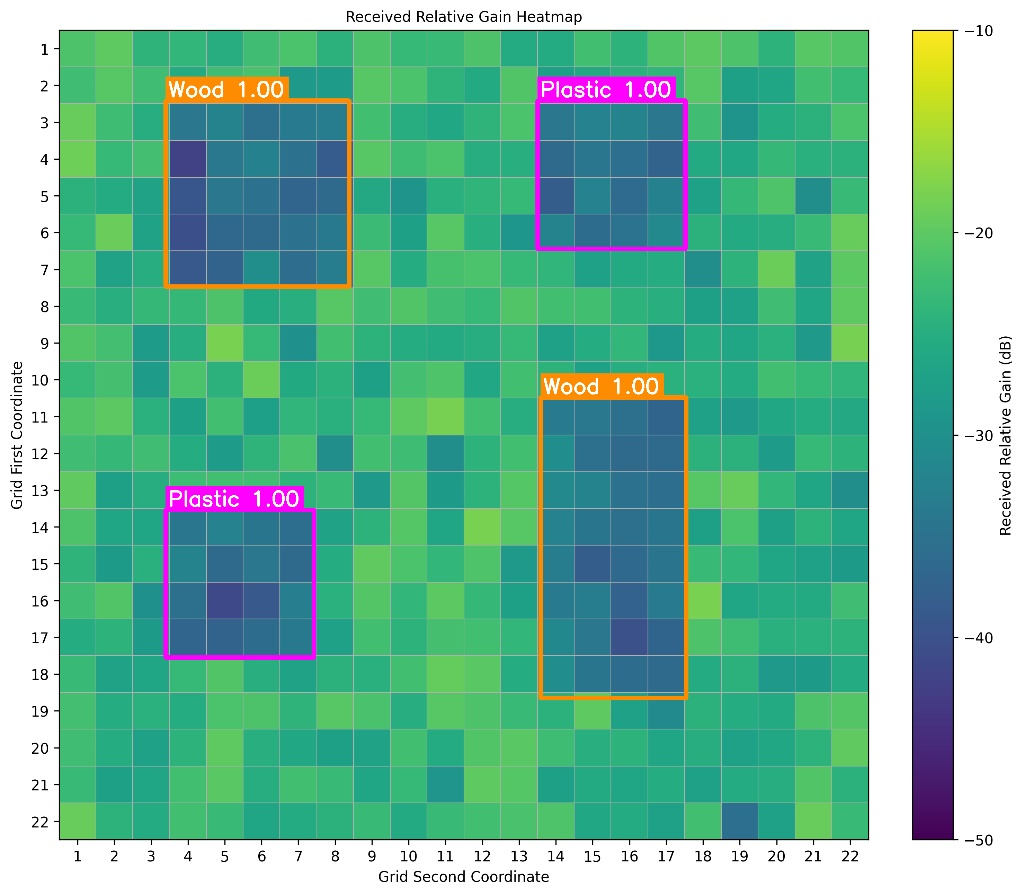}
    \caption{\color{blue}A detection example after sim-to-real mixed training.}
    \label{fig:sim-to-real}
\end{figure}

\noindent\textbf{Results.}
Synthetic-only training localizes only 3/8 objects on
measured maps, indicating a substantial sim-to-real gap.
Real-only training improves localization to 5/8 objects,
but correctly identifies the materials of only 2/8.
In contrast, joint training achieves 8/8 correct localization
and material identification on the measured test maps, Fig. \ref{fig:sim-to-real} shows an example of detection on real-data map.

The jointly trained model also generalizes well to unseen
synthetic scenarios, correctly detecting 580/584 objects
(99.3\%), with all four objects correctly identified in
142/146 maps (97.3\%). Detection performance stabilizes
within 700 training steps and remains unchanged through
3,000 steps. Varying the confidence threshold from
0.3 to 0.9 also leaves the results unchanged.

These findings suggest that combining real-data measurements
with synthetic data can improve real-world detection while
maintaining strong generalization within the synthetic domain.

\noindent\textbf{Observation 5:}
\textit{Real and synthetic data could be complementary:
neither alone achieves the detection performance of their
combination. Co-training bridges the sim-to-real gap
while preserving generalization to unseen synthetic scenarios.}
}

\subsection{Comparative Analysis with State-of-the-Art} 
To contextualize the computational efficiency of \name, we benchmark it against state-of-the-art vision-based and RF-based detection frameworks \cite{wang2023, zhou2024yolo, AmatareICC2025}. As detailed in Table~\ref{tab:sota_comparision}, \name~provides an optimal trade-off between latency, privacy, and spatial flexibility. With a highly compact parameter footprint ($135$M) and an inference latency of $8.2$ ms on an A100 GPU, {\color{blue} it is compatible with MEC-server-side deployment.} Crucially, unlike RF-Vision \cite{AmatareICC2025}, which relies on centralized static transmitters and retrofitted closed-source image models, \name~supports arbitrary, mobile transmitter placements while preserving user privacy. This explicitly validates Supporting Contribution 1.

\begin{table*}[t!]
\color{blue}
\centering
\caption{\color{blue}Performance Comparison with State-of-the-Art Detection Frameworks.}
\label{tab:sota_comparision}
\renewcommand{\arraystretch}{1.2}
\resizebox{0.7\textwidth}{!}{
\begin{tabular}{l c c c c c}
\toprule
\textbf{Framework} & \textbf{Architecture} & \textbf{Hardware} & \textbf{Inference} & \textbf{Privacy} & \textbf{Tx Placement} \\
& & & \textbf{Time (ms)} & \textbf{Risk} & \textbf{Constraint} \\
\midrule
Wang {\em et al.} \cite{wang2023} & YOLOv7 & RTX 4090 & 5.4 & High (Visual) & N/A (Optical) \\
Zhou \cite{zhou2024yolo} & YOLO-NL & Ryzen 7 & 10.4 & High (Visual) & N/A (Optical) \\
RF-Vision \cite{AmatareICC2025} & YOLOv11n & A100 & 2.2 & Low (RF) & Fixed (Center) \\
\textbf{\name~(Ours)} & \textbf{ViT-DETR} & \textbf{A100} & 8.2 & \textbf{Low (RF)} & \textbf{Arbitrary/Mobile} \\
\bottomrule
\end{tabular}}
\end{table*}


{\color{blue}\subsection{Practical Deployment and Co-existence with ISAC Infrastructure}
\label{subsec:isac_infrastructure}

A critical practical consideration is the acquisition of the dense RF propagation maps required as input for the model. \name does \textit{not} require a dedicated roaming robot to manually survey the floor at inference time, nor does it necessitate a dense physical grid of hardware receivers. Instead, it leverages pre-existing Integrated Sensing and Communications (ISAC) infrastructure. 

In typical modern indoor deployments, RF transmitters (e.g., Wi-Fi access points or 6G mobile hubs) continuously monitor the environment. The dense RF propagation map is synthesized at the Mobile Edge Computing (MEC) server using the Digital Twin, which is continuously calibrated using sparse, real-time channel measurements obtained from these existing APs and active IoT nodes. \name utilizes this MEC-generated map to perform rapid material characterization. This allows facility operators to achieve high-fidelity environmental perception with zero additional hardware overhead, effectively providing "sensing-as-a-service" to passive robotic agents on the floor.}

{\color{blue}\subsection{Limitations}
\label{sec:limitations}
We explicitly note limitations of the present study.

\noindent\textbf{Static scenes without human blockage.} Every propagation map is
synthesized for a static scene: object positions and materials are
frozen within a block of 500 samples, and no humans or moving scatterers
are present. Transient blockage by people, doors, or carts alters the
multipath structure on time scales of seconds and is a known failure
mode for propagation-map perception. PRISM's robustness to such
spatio-temporal dynamics is untested.

\noindent\textbf{No edge-hardware timing.} The reported inference latency
(8.2\,ms, Table~\ref{tab:sota_comparision}) and adaptation cost (42\,s, 3.8\,GB
VRAM) are measured on an NVIDIA A100. We have not profiled PRISM on
representative MEC or embedded accelerators (e.g., Jetson-class GPUs or
CPU-only edge servers), where a 135M-parameter ViT-DETR may require
quantization or distillation to meet real-time budgets. Our claim of
edge suitability is thus based on parameter count and A100 latency
rather than on-device measurements.

\noindent\textbf{Dense-input assumption.} PRISM
consumes a dense propagation map sampled at roughly 1\,m over the entire
floor. In the ISAC deployment sketched in Section~\ref{subsec:isac_infrastructure}, this
map is synthesized at the MEC server from a digital twin calibrated with
sparse channel measurements. We have not characterized how detection
accuracy degrades as the number or placement of physical measurement
points decreases, nor how errors in the upstream radio-map estimator
propagate to the detector.

\noindent\textbf{Limited Sim-to-Real Co-Training Evaluation Settings.}
The measured Sim-to-Real Co-training evaluation is limited to two test maps from
a single room. Further evaluation across different
rooms and object configurations is needed to establish broader
real-world generalization.}

%% file: Sections/8-conclusion.tex
\section{Conclusions}
\label{sec:conclusions}

In this paper, we presented \name, a lightweight, edge-native, and privacy-preserving framework for RF-based indoor material characterization. \name~is specifically architected to overcome three critical barriers prevalent in existing RF sensing methodologies: transmitter location dependency, a lack of cross-scenario generalization, and the rigid closed-set material assumption. Built upon a vision-centric ViT backbone paired with a DETR-style set-prediction head, our framework natively captures the global electromagnetic dependencies inherent in diffuse RF multipath propagation.  Trained on a deterministic, ray-traced DT dataset, \name~demonstrates exceptional spatial robustness. Through a novel few-shot support set adaptation strategy utilizing only $25$ samples, the model achieved an overall F1 score of $0.9822$ on a severely out-of-distribution benchmark. Furthermore, it maintained high spatial localization fidelity (mean IoU of $0.86$) even under challenging, near-grazing LoS conditions. To address the unpredictability of open-world deployments, we pioneered an outlier exposure strategy operating in suppression mode. This approach successfully achieved a $100\%$ anomaly rejection rate for unknown materials without compromising the detection accuracy of known environmental features, decisively outperforming post-hoc thresholding techniques. {\color{blue}
We further validate \name~on real-world RF measurements,
where joint training with measured and DT-synthesized maps
correctly localizes and identifies all eight objects in
two held-out real maps, outperforming either data source
alone. The jointly trained model also maintains 99.3\%
detection accuracy on unseen synthetic scenarios,
demonstrating the complementary benefits of both data sources.
}
{\color{blue}Future research will focus on four primary trajectories: (1) expanding the deterministic data synthesis pipeline to encompass sub-THz frequency bands relevant to 6G ISAC; (2) further mitigating the sim-to-real domain gap through expanded empirical RF calibration deployments; (3) advancing the anomaly suppression mechanism to output explicit bounding boxes for unknown materials; and (4) extending the framework to model spatio-temporal dynamics, specifically evaluating the robustness of \name~against transient human blockage and moving scatterers in highly populated environments.}